\documentclass[twocolumn,prb,aps,showpacs,superscriptaddress,floatfix]{revtex4}

\usepackage{graphicx}
\usepackage{dcolumn}
\usepackage{bm}

\begin{document}

\title{Quantum critical behavior of the one-dimensional ionic Hubbard model}

\author{S.R.\ Manmana}
\affiliation{Institut f\"ur Theoretische Physik III, Universit\"at Stuttgart,
  Pfaffenwaldring 57, D-70550 Stuttgart, Germany}
\affiliation{Institut f\"ur Theoretische Physik, Universit\"at
  G\"ottingen,  Tammannstr.\ 1,   
  D-37077 G\"ottingen, Germany}
\author{V.\ Meden}
\affiliation{Institut f\"ur Theoretische Physik, Universit\"at
  G\"ottingen,  Tammannstr.\ 1, 
  D-37077 G\"ottingen, Germany}
\author{R.M.\ Noack} 
\affiliation{Institut f\"ur Theoretische Physik III, Universit\"at Stuttgart,
  Pfaffenwaldring 57, D-70550 Stuttgart, Germany}
\author{K.\ Sch\"onhammer}
\affiliation{Institut f\"ur Theoretische Physik, Universit\"at
  G\"ottingen, Tammannstr.\ 1, 
  D-37077 G\"ottingen, Germany}

\date{Preliminary version of \today}


\begin{abstract}
We study the zero-temperature phase diagram of 
the half-filled one-dimensional ionic Hubbard model. 
This model is 
governed by the interplay of the on-site Coulomb repulsion and 
an alternating one-particle potential. 
Various many-body energy 
gaps, the charge-density-wave and bond-order parameters,
the electric as well as the bond-order 
susceptibilities, 
and the density-density correlation function
are calculated using the density-matrix
renormalization group method. 
In order to obtain a comprehensive picture, we investigate systems
with open as well as periodic boundary 
conditions and study the physical properties in different sectors of the
phase diagram. 
A careful finite-size scaling analysis 
leads to results which give strong evidence in favor of a scenario 
with two quantum critical points
and an intermediate spontaneously dimerized phase.
Our results indicate that the phase transitions are continuous.
Using a scaling ansatz we are able to read off critical exponents at
the first critical point. 
In contrast to a bosonization approach, we do 
not find Ising critical
exponents.  
We show that the low-energy physics of the 
strong coupling phase can only partly be
understood in terms of the strong coupling behavior of the ordinary
Hubbard model. 

\end{abstract}

\maketitle

\section{Introduction}
\label{sec:intro}

\subsection{Motivation}
\label{subsec:motivation}

Theoretical studies of the ionic Hubbard model (IHM) date back as far as 
the early seventies 
(see Ref.\ \onlinecite{Nagaosa} and references therein). 
The model consists of the usual Hubbard model with on-site Coulomb 
repulsion $U$ supplemented by an alternating one-particle
potential of strength $\delta$. 
It has been used to study the neutral to ionic transition in organic
charge-transfer 
salts\cite{Hubbard,Nagaosa} and to understand the ferroelectric
transition in perovskite materials.\cite{Egami}  
Based on results obtained 
from numerical\cite{Soos,Resta} and approximate
methods,\cite{Strebel,Ortiz} it was generally believed that 
at temperature $T=0$ and for fixed $\delta$ a single phase transition 
can be found if $U$ is varied. This 
quantum phase transition 
was also interpreted as an
insulator-insulator transition from a band insulator ($U \ll \delta$)
to a correlated insulator ($U \gg \delta$). 
In the present paper, we discuss in detail how this transition occurs.

In 1999, Fabrizio, Gogolin, and Nersesyan used bosonization to 
derive a field-theoretical model which they argued to be the effective 
low-energy model of the one-dimensional IHM.\cite{Fabrizio} 
Surprisingly, the authors found, using various approximations, that
the field-theoretical model displays {\it two}
quantum critical points as $U$ is varied for fixed $\delta$. 
For $ U < U_{\text{c1}}$ the system is a band 
insulator 
(with finite bosonic spin and charge gaps), as expected from
general arguments.
At the first transition point $U_{\text{c1}}$, they found Ising critical
behavior as well as metallic behavior in the sense that the gap to the
bosonic charge modes goes to zero at the critical point only.
In the intermediate regime,  $U_{\text{c1}} < U < U_{\text{c2}}$, a 
spontaneously dimerized insulator phase 
(in which the bosonic spin and charge gaps are finite) 
with finite bond order (BO) parameter was found. 
The authors argued that the system goes over into a correlated
insulator phase 
(in which the bosonic charge gap is finite) 
with vanishing 
bond order and bosonic spin gap at a second
critical point $U_{\text{c2}}$ which is of Kosterlitz-Thouless 
(KT) type. 

Several groups have attempted to verify this phase diagram for the IHM 
using mainly numerical methods. 
Variational and Green's function Quantum Monte Carlo (QMC) data
obtained for the BO parameter, the electric polarization, and the 
localization length were interpreted in favor of a scenario with a 
single critical point $U_{\text{c}}$ and finite BO for 
$U>U_{\text{c}}$.\cite{Wilkens} 
In a different calculation using auxiliary field QMC, 
data for the one-particle spectral weight were argued to show two
critical points with an intermediate metallic phase.\cite{Refolio} 
Exact diagonalization studies of the Berry phase\cite{Torio} and
energy gaps\cite{Gidopoulos,Pati,Brune} have been interpreted as
favoring one critical point\cite{Pati} or two points;\cite{Torio} 
in two investigations this issue was left unresolved.\cite{Gidopoulos,Brune} 
Several density-matrix renormalization group (DMRG) studies have
been performed focusing on different energy
gaps,
the localization length,
the BO parameter, the BO correlation function,
different distribution functions,
and the optical conductivity.
\cite{Takada,Qin,Brune,Zhang}
Some of the results have been
interpreted to be consistent with a two-critical-point
scenario.\cite{Takada,Qin,Zhang} 
In Ref.\ \onlinecite{Brune} the signature of only one
phase transition was found and 
the possible existence of a second
transition was left undetermined. 
The phase diagram of the IHM has also been studied 
using  approximate methods such as the self-consistent mean-field
approximation\cite{Caprara,Gupta,Gros}, the slave-boson
approximation,\cite{Caprara} and a real space renormalization group 
method.\cite{Gupta} 
Although these studies led to interesting
insights, the validity of the approximations in the vicinity of the 
critical region can be questioned on general grounds; therefore, we do
not focus on these approaches any further here.  
The present situation can be summarized as being highly
controversial. 

Here we refrain from giving a detailed 
discussion of the merits and shortcomings of the various numerical 
methods used and the possible problems in interpretation of numerical results 
in the literature. 
Instead, we present a detailed study of the 
$T=0$ phase diagram of the one-dimensional IHM 
mainly
based on 
DMRG calculations on systems with both open and periodic boundary conditions
(OBC's and PBC's).

We have calculated a number of different many-body energy gaps, 
including the spin gap, the one-particle gap
(the energy difference of the ground states with $N+1$, $N$, and $N-1$ 
electrons), and the gaps to the first (``exciton'') and second
excited states.
A definition of the gaps is given in Sec.\ \ref{subsec:def}.
Our results explicitly show that different gaps 
associated with charge degrees of freedom
do not coincide in the thermodynamic limit,
although they are often believed to in the literature (see also 
Refs.\ \onlinecite{Qin} and \onlinecite{Brune}). 
Our data show that the exciton gap vanishes at a coupling which
depends on $\delta$ and which we define as $U_{\text{c1}}$. At this
critical point the spin gap remains finite. The spin gap vanishes at 
a second critical coupling, which defines our $U_{\text{c2}}$.

In addition to the energy gaps, we have determined the BO parameter
and susceptibility as well as the charge-density-wave (CDW)
order parameter.
Since the single-site translational symmetry is explicitly 
broken due to the alternating potential, 
we will avoid using the term ``order parameter'' in describing the CDW
order and instead use the term ``ionicity'' 
to refer to the difference
in occupancy between sites on the two sublattices $\langle n_A -n_B \rangle$.
We find that the ionicity is continuous and non-vanishing for all values
of the interaction strength.

From the finite-size scaling of the BO parameter, we find a parameter
regime with a non-vanishing dimerization 
starting at $U_{\rm c1}$ and ending at $U_{\rm c2}$. 
We find that the transitions at both
critical points are continuous.
The BO susceptibility shows one isolated divergence at $U_{\text{c1}}$
separated from a region of divergence starting at $U_{\text{c2}}$.

We have also investigated the electric 
susceptibility, which is finite in the thermodynamic limit for
$U<U_{\text{c1}}$ and diverges at the lower transition point 
$U_{\text{c1}}$.
For $U>U_{\text{c1}}$,
the behavior is less clear: 
there seems to be a weak divergence with system size near 
$U_{\text{c2}}$ and for $U > U_{\text{c2}}$.
This behavior is consistent with that of the density-density
correlation function, which decays exponentially as expected 
in a band insulator phase for $U<U_{\text{c1}}$, but 
surprisingly decays as a power law with an exponent
between 3 and 3.5 in the strong coupling regime, $U>U_{\text{c2}}$.  

Using a scaling ansatz for the BO and the electric 
susceptibility we can determine the critical exponents at 
$U_{\text{c1}}$. 
In contrast to the bosonization approach\cite{Fabrizio}, 
we obtain critical exponents different from those of the
two-dimensional Ising model. 

For (almost) all observables, we find that a careful finite-size
scaling analysis is crucial to obtain reliable results in the
thermodynamic limit.
Furthermore, since it is necessary to distinguish between fairly small, but 
finite, gaps and order parameters and vanishing ones, a
detailed understanding of the accuracy of the DMRG data is essential.

In order to obtain a comprehensive picture of the ground-state phase
diagram, we have studied the
different phases (as a function of $U$) 
for different $\delta$'s which cover a wide range of the parameter
space.
We also consider the limit of large Coulomb
repulsion $U \to \infty$ (for fixed $\delta$ and hopping matrix
element $t$) and show that 
some aspects of
the physics of the model 
in this limit can be understood in terms of an effective Heisenberg 
model, as has been suggested earlier\cite{Nagaosa} but has recently
been questioned.\cite{Brune}
As a result of our investigations, we are able to resolve many 
of the controversial issues and present strong indications in favor 
of a scenario with
two quantum critical  points.
At the appropriate points in the paper, we will
briefly comment on the relationship of our results with the ones 
obtained in earlier publications.

The remainder of the paper is organized as follows. 
In Sec.\ \ref{subsec:model}, we introduce the model and
discuss the  limits in which it can be 
treated exactly. 
In Sec.\ \ref{subsec:DMRG}, we discuss the  details of our DMRG procedure. 
In Sec.\ \ref{sec:gaps}, our finite-size and extrapolated data
for the energy gaps are discussed. 
In Sec.\ \ref{sec:cdw}, we present our results for 
the ionicity and show that in the large-$U$ limit they are consistent
with analytical results obtained by mapping the IHM to an effective
Heisenberg model. 
The BO parameter and the related susceptibility are investigated in
Sec.\ \ref{subsec:bow}. 
We present results for the electric
susceptibility 
and the density-density correlation function
in Sec.\ \ref{subsec:polsus}. 
In the numerical calculations of Secs.\ \ref{sec:cdw} to 
\ref{sec:orderparams} we use OBC's, for which the DMRG algorithm
performs best. 
To complete our DMRG study in Sec.\ \ref{sec:pbc}, we present
results for the energy gaps calculated for PBC's and summarize our
findings in Sec.\ \ref{sec:sum}.

\subsection{Model and exactly solvable limits}
\label{subsec:model}

The one-dimensional IHM is given by the Hamiltonian 
\begin{eqnarray}
H & = & -t \sum_{j,\sigma} \left( c_{j\sigma}^{\dag} c_{j+1\sigma}^{} +
  \mbox{h.c.}  \right) + U \sum_{j} n_{j\uparrow} n_{j \downarrow}
\nonumber \\ && + \frac{\delta}{2} \sum_{j,\sigma} (-1)^j n_{j\sigma}
\; , 
\label{Hamiltonian}
\end{eqnarray}
where $ c_{j\sigma}^{}$ ($ c_{j\sigma}^{\dag}$) destroys (creates) an
electron with spin $\sigma$ on lattice site $j$ and 
$n_{j\sigma} = c_{j\sigma}^{\dag}c_{j\sigma}^{}$. 
We set the lattice constant equal
to 1 and denote the number of lattice sites by $L$. 
Here we study the properties of the half-filled system 
with $N=L$ electrons.  

The system corresponds to the usual Hubbard model 
with an additional local alternating potential.
It is useful to consider various limiting cases in order to gain
insight into possible phases and phase transitions.
For $U=0$ and $\delta >0$, the model describes a conventional 
band insulator with a band gap $\delta$. 
Since the  alternating one-particle potential explicitly breaks the 
one-site translational symmetry, the ground state has finite ionicity. 

The one-dimensional half-filled Hubbard model 
without the alternating potential ($\delta=0$) and with $U > 0$ describes a
correlated insulator with vanishing  
spin gap $\Delta_{\text{S}}^{\rm HM}(U)$ and critical spin-spin and bond-bond
correlation functions.\cite{Voit}
All gaps associated with the charge degrees of freedom, such as the 
one-particle gap $\Delta_1^{\rm HM}(U)$, are finite.\cite{Florian} 
(The gaps discussed here are defined in Sec.\ \ref{subsec:def}.)
The ionicity and the dimerization are zero
for all values of $U$.
These two limiting cases suggest that the system will be in two
qualitatively different phases in the limits $U \ll \delta$ and 
$U \gg \delta$. 

In the atomic limit, $t=0$, and for $0< U < \delta$, every second site
of the lattice with on-site energy $-\delta/2$ (A sites) is occupied
by two electrons while the sites with energy $\delta/2$ (B sites) are
empty.  The energy difference between the ground state and the highly
degenerate first excited state is $\delta-U$.  For $U > \delta$, both
the A and B sites are occupied by one electron and the energy gap is
$U - \delta$.  Thus for $t=0$ a single critical point
$U_{\text{c}}(\delta)=\delta$ with vanishing excitation gap can be found.  One
expects similar critical behavior with at least one critical point to
persist for the full problem with finite $t$.

To describe the physics of the IHM in the limit $U \gg t,\delta$, an 
effective Heisenberg Hamiltonian 
\begin{eqnarray}
H_{\rm HB} = J \sum_{j} \left( {\mathbf S}_{j} \cdot {\mathbf S}_{j+1}
- \frac{1}{4}\right) \; , \;\;\; J= \frac{4 t^2 U}{U^2 - \delta^2}   
\label{effectiveHeisenberg}   
\end{eqnarray}
was derived in Ref.\ \onlinecite{Nagaosa} 
analogously to the strong-coupling perturbation
expansion of the usual Hubbard model. 
It has recently been pointed out
that this strong-coupling mapping 
does not take into account
an explicitly broken
one-site translational symmetry.\cite{Brune}
However, it was shown in Ref.\ \onlinecite{Nagaosa} that the
strong-coupling expansion preserves the one-site translation symmetry
in the effective spin Hamiltonian to {\it all} orders in the
strong-coupling expansion.
In addition, the ionicity can be derived directly from the effective
spin Hamiltonian as follows.
The symmetry of the Hamiltonian [Eq.\ (\ref{Hamiltonian})] implies that 
after taking the thermodynamic limit, 
$n_{j\sigma} = n_{j+2\sigma}$ for $\sigma= \uparrow,\downarrow$ and 
all $j$. Using 
the Hamiltonian
Eq.\ (\ref{Hamiltonian}) and 
the Hellman-Feynman theorem, the ionicity
\begin{equation}
  \langle n_A -n_B \rangle = - \frac{2}{L} \sum_{j,\sigma} 
  (-1)^{j} \langle n_{j\sigma} \rangle 
  \label{CDWdef}
\end{equation} 
can be determined via
\begin{eqnarray}
\langle n_A -n_B \rangle = - \frac{4}{L} \left\langle   \frac{\partial
    H}{\partial \delta} \right\rangle  = - \frac{4}{L} \frac{\partial
    E_0}{\partial \delta} \; . 
\label{nAnBHamiltonian}
\end{eqnarray}
The ground-state energy $E_0$ of the effective
Heisenberg model [Eq.\ (\ref{effectiveHeisenberg})] is known 
analytically\cite{HBenergy} and, in terms of $U$ and $\delta$, is given by 
\begin{eqnarray}
E_{0}^{\text{HB}} = L \frac{4  U t^2}{U^2 - \delta^2} \left( \ln 2 - \frac{1}{4}
  \right) 
\label{HBenergy}
\end{eqnarray}
in the thermodynamic limit.
In the limit $ U \gg \delta$, we can thus derive an analytic
expression for the ionicity 
\begin{eqnarray}
\langle n_A -n_B \rangle = 32 \ln 2 \; \frac{U \delta t^2}{\left(U^2 - 
\delta^2\right)^2} \; .
\label{nAnBanalytical}
\end{eqnarray}
It implies that for any $U < \infty$, the ionicity of the
IHM is nonzero and for large $U$ vanishes as $1/U^3$. 
Since CDW order is explicitly favored by the Hamiltonian, it is not
surprising that the ionicity is non-vanishing for
all finite $U$.
As will be shown in Sec.\ \ref{sec:cdw},
this expression  shows excellent agreement with our DMRG data for the IHM.
This gives us confidence that the effective Heisenberg model indeed 
gives correctly at least certain aspects of the
low-energy physics. 
Since the Heisenberg model [Eq.\ (\ref{effectiveHeisenberg})] has 
a vanishing spin gap,\cite{HBzerospingap} the mapping 
suggests that the spin gap also vanishes in the large-$U$ limit of the 
IHM.

Although the alternating potential breaks the one-site translational
symmetry explicitly, the model remains invariant to a
translation by two lattice sites. 
This leads to a
site-inversion symmetry for closed-chain geometries with 
periodic or antiperiodic boundary
conditions, a symmetry which is not present for OBC's.
As pointed out in Ref.\ \onlinecite{Gidopoulos},
the ground state of the effective Heisenberg model 
with periodic boundary conditions for systems with $4n$ lattice sites or
antiperiodic boundary conditions for systems with $4n+2$ sites
has a parity eigenvalue of $-1$ whereas the ground state
for $U=0$ has a parity eigenvalue of $+1$.
This suggests that the IHM undergoes at least one phase
transition point with increasing $U$ for fixed $\delta$. 
This level crossing will be replaced by level repulsion and
approximate symmetries for other boundary conditions.
In the thermodynamic limit, the effect of the boundaries will
disappear and the level repulsion becomes vanishingly small.
It is important to point out, however, that a level crossing on small
finite systems does not necessarily lead to a first-order transition
in the thermodynamic limit; careful finite-size scaling must be
carried out in order to determine the critical behavior.

From these considerations, one expects to find at least one quantum
phase transition from a phase with physical properties similar to
those of a non-interacting band insulator to a phase with properties 
similar to those of the strong coupling phase of the ordinary Hubbard
model. 
However, the details of the transition and the physical properties 
of the different phases remain unclear
from these arguments. 
Furthermore, the behavior
of the BO parameter in the critical region cannot be
estimated from these simple limiting cases. 
Therefore, a detailed and careful calculation of the characterizing
gaps and order parameters is necessary. 
Since no direct analytic approach is known to be able to treat the
parameter values in the critical regime, we restrict ourselves to
numerical calculations using the DMRG method with the details
described in the next section.

In the following, we measure energies in units of the hopping matrix
element $t$, i.e., set $t=1$. 
In order to be able to cover a significant part of the parameter space,
we have carried out calculations with $\delta = 1$, $\delta=4$, and $\delta
= 20$ for weak interaction values $U \ll \delta$, for strong coupling $U \gg
\delta$ and in the intermediate critical regime 
$U \approx \delta$.
For the sake of compactness, we will 
mostly 
focus on $\delta=20$ when
presenting results that are generic to all three
$\delta$-regimes.

\subsection{DMRG method}
\label{subsec:DMRG}

We have carried out our calculations using the finite-system DMRG algorithm. 
Our investigation focuses on the ground-state properties for systems
with OBC's, i.e., we have performed DMRG runs mostly with OBC's and one target
state, the case in 
which the DMRG algorithm is most efficient.
In order to perform the demanding  finite-size scaling  
necessary, we have performed calculations for systems with up to 
$L = 768$ sites, much larger than in an earlier work.\cite{Schoenhammer}

In order to investigate the low-lying excitations, we have also
performed calculations targeting up to three states
simultaneously on systems with OBC's. 
These numerically more demanding calculations were carried out for systems
with up to $L = 256$ sites for three target states and with up to 
$L = 450$ sites for two target states.
In order to compare with exact diagonalization calculations and to
extend its finite-size scaling to larger systems,
we have performed calculations for PBC's with up to $L = 64$ sites
and one to three target states. 
In this case, the maximum system size is limited by the
relatively poor convergence.

The DMRG calculations for OBC's with one target state were carried out
performing up to six finite-system sweeps keeping up to $m = 800$ states. 
For more multiple states and for PBC's up to 12 sweeps  were performed,
keeping up to $m=900$ states. 
In order to test the convergence of the DMRG runs, the sum of the
discarded density-matrix eigenvalues and the convergence of the 
ground-state energy were monitored.
For OBC's, 
the discarded weight is of order $10^{-6}$
in the worst case 
and the ground-state energy 
is converged to an absolute error of $10^{-3}$ 
but in most cases the
absolute error is $10^{-5}$ or better.
This accuracy in both the energy 
and the discarded weight
gives us confidence that the wave function
is also well-converged and that local quantities are quite accurate.

For PBC's, the discarded weight is of the order $10^{-5}$ 
in the worst case 
and the convergence of the
ground-state energy for most runs is up to
an absolute error of $10^{-3}$
or better, but for extreme cases 
such as
$L=64$ and three target
states for parameter values near the phase transition points, the
convergence in the energy is sometimes 
reduced to an absolute error of only $10^{-1}$.
However, we believe that this accuracy is high enough for the
purposes of the discussion in section \ref{sec:pbc}.

In general,
we find that our data are  
sufficiently accurate so that extrapolation 
in the number of states $m$ kept in the DMRG procedure does
not bring about significant improvement in the results
(at least for OBC's).
Details of the extrapolations and error estimates for particular calculated
quantities are given in the corresponding sections. 

\section{Energy gaps}
\label{sec:gaps}

One important way to characterize the different phases of the
IHM are the energy differences between many-body eigenstates. 
Gaps to 
excited states 
can be used to characterize phases by making
contact with the gaps obtained in bosonization calculations and also
form the basis for experimentally measurable excitation gaps, found,
for example, in inelastic neutron scattering, optical conductivity, or
photoemission experiments.
In addition to the gaps themselves, however, 
matrix elements between ground and excited states as well as the
density of excited states are important in forming the full
experimentally relevant dynamical
quantities.
An example is the matrix element of the current operator that comes
into calculations of the optical conductivity.
We have investigated the behavior of the matrix elements for the
dynamical spin and charge structure factors and for the optical
conductivity using exact diagonalization on systems with both PBC's
and OBC's.

In the following,
we present DMRG calculations of the gaps to first and
second excited states,  
the spin gap, and the
one-particle gap in which a careful finite-size scaling on systems of
up to 512 sites is carried out.
As we shall see, 
this 
is necessary in
order to resolve the behavior of the gaps in the transition regime and
to distinguish between scenarios with one or two critical points.

\subsection{Definition of the gaps}
\label{subsec:def}

In this section,
we study excitations between a non-degenerate $S=0$ ground
state and various excited states.
In the numerical calculations, we have found that for OBC's
the ground state is non-degenerate with total spin $S=0$ for all
parameter values studied here.
We define 
the exciton gap
\begin{eqnarray}
\Delta_{\text{E}} =  E_1(N,S) - E_0(N,S=0)
\label{Egap}
\end{eqnarray}
as the gap to the first excited state in the sector with the same
particle number $N$ and with $S_z=0$, where $S_z$ is the $z$-component of
the total spin.
We also calculate the expectation value of the total spin operator
$\langle {\bf S}^2 \rangle$ so that $S$ is known.

The spin gap is defined
as 
the energy difference between the ground state and the 
lowest lying energy eigenstate in the $S=1$ subspace 
\begin{eqnarray}
\Delta_{\text{S}} & = & E_0(N,S=1) - E_0(N,S=0) \; .
\label{Sgap}
\end{eqnarray}
When the first excited state $E_1(N,S)$ in the $S_z=0$ subspace is a
spin triplet with $S=1$, $\Delta_{\text{S}}  = \Delta_{\text{E}}$. 
Within the DMRG, this gap can be calculated by determining the
ground-state energies in different $S_z$
subspaces in two different DMRG runs.

If $\Delta_{\text{E}} < \Delta_{\text{S}}$,
we {\it call} the lowest excitation a charge
excitation.
In fact, exact diagonalization calculations for system with PBC's
suggest that the gap $\Delta_{\text{E}}$ corresponds to the gap in the optical
conductivity.\cite{Brune}
We have carried out additional exact diagonalization calculations that
show that the corresponding matrix elements of the current operator
are also nonzero for OBC's.
We therefore expect that $\Delta_{\text{E}}$ (for excitations with $S=0$ and
when $\Delta_{\text{E}} < \Delta_{\text{S}}$)
corresponds to the optical gap in the 
thermodynamic limit.\cite{diploma}
To obtain a deeper understanding of the excitation spectrum in the
critical region, we also calculate the 
gap to the second excited state
\begin{eqnarray}
\Delta_{\text{SE}} = E_2(N,S) - E_0(N,S=0)
\label{2gap}
\end{eqnarray} 
for selected parameters.

In the literature, gaps to excitations 
which can be classified as charge excitations
are often calculated 
by taking differences between ground-state energies in sectors with 
different numbers of particles
(this gap is commonly called the ``charge gap'').
In particular, one can define a $p$-particle gap
\begin{eqnarray}
\Delta_p & = & \left [ E_0(N+p,S^z_{\text{min}}) +
  E_0(N-p,S^z_{\text{min}}) \right . \nonumber \\
& & \left . - 2 E_0(N,S=0) \right ] /p  
\label{1gap}
\end{eqnarray} 
which is essentially the difference in chemical potential for adding and
subtracting $p$ particles.
The spin $S^z_{\text{min}}$ is the minimal value, 1/2 or 0 for $p$ odd
and even, respectively.
Either the one particle gap $\Delta_1$ or
the two particle gap $\Delta_2$ are commonly used.
The calculation of $\Delta_1$ or $\Delta_2$ is numerically less
demanding than that of $\Delta_{\text{E}}$ since it is sufficient to calculate
the ground-state energies in the subspaces with the corresponding
particle numbers.
However, since these gaps 
involve changing the particle number and, for $p=1$, the spin quantum
number, 
it is not a priori clear if they can be used to characterize
possible phase transition points of the $N$-particle system. 
In many cases of interest, the difference between $\Delta_1$,
$\Delta_2$, and $\Delta_{\text{E}}$ vanishes for $L \to \infty$, but in other
systems (an example is the Hubbard chain with an attractive
interaction), their behavior differs.
As we shall see, $\Delta_1$ and $\Delta_{\text{E}}$ do behave differently near
$U_{\text{c1}}$.
In this work we
focus our investigation on $\Delta_1$.
We have also calculated $\Delta_2$ and find that it behaves similarly
to $\Delta_1$, although it generally takes on slightly larger values
for finite systems.

\begin{figure*}[t]
\includegraphics[width=0.46 \textwidth]
%{psfiles/delta20/excitongap/exzitongap_delta20_obc.eps}
{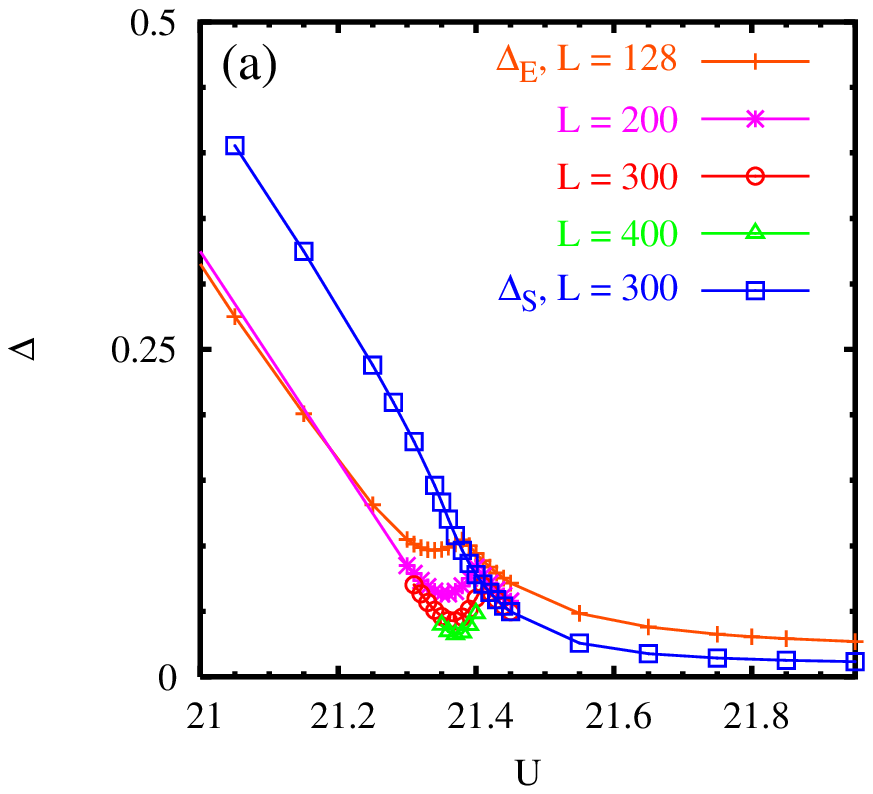}
\includegraphics[width=0.45 \textwidth]
%{psfiles/delta20/excitongap/sitrigap_L128_fine_delta20_obc.eps}
{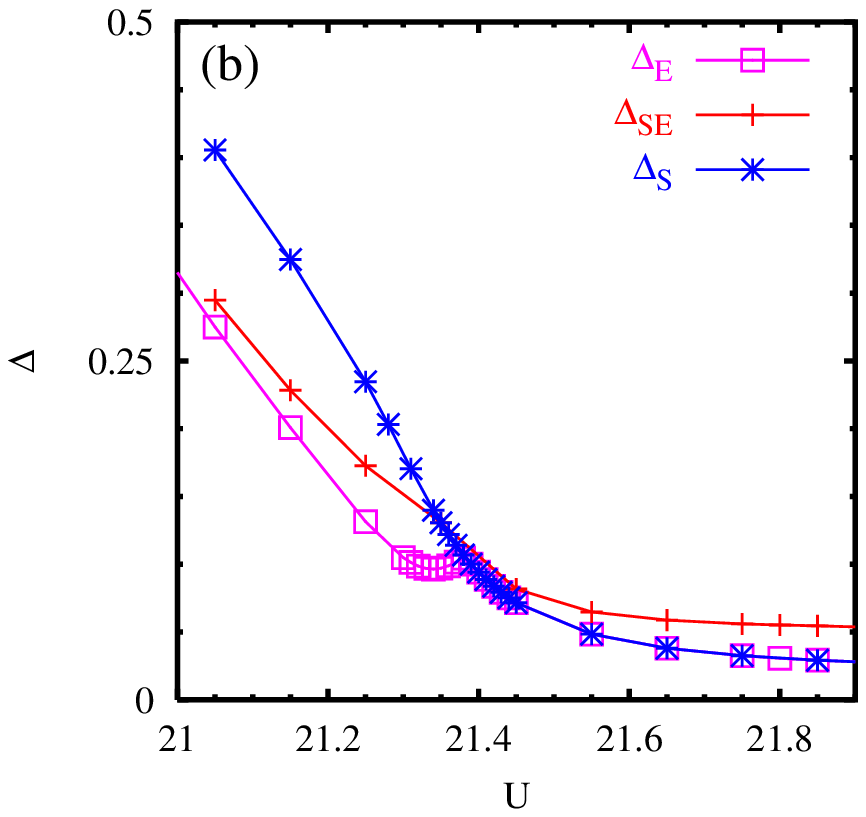}
\caption{(a) The exciton gap $\Delta_{\text{E}}$ for finite system sizes $L$
  and $\delta=20$.
  The spin gap $\Delta_{\text{S}}$ for $L=300$ is also shown for comparison.
  (b) The exciton gap, the spin gap $\Delta_{\text{S}}$ and the gap to the
  second excited state $\Delta_{\text{SE}}$ for $L=128$.
}
\label{fig:exgap_all}
\end{figure*}

Gaps are also used to characterize the phase diagram
within the bosonization approach.\cite{Fabrizio,Brune}  
It is generally believed that the bosonic charge gap defined there 
can be identified with the gap to the first excited state with spin 
quantum number $S=0$ (i.e.\ the exciton gap $\Delta_{\text{E}}$ 
[Eq.\ (\ref{Egap})] as long as $\Delta_{\text{E}} < \Delta_{\text{S}}$) 
and the bosonic spin gap with $\Delta_{\text{S}}$ 
[Eq.\ (\ref{Sgap})], although a formal proof is missing.

Based on $\Delta_{\text{E}}$, $\Delta_{\text{S}}$, and 
$\Delta_{\text{SE}}$ and the very limited
knowledge on matrix elements due to the small system sizes available 
to exact diagonalization, 
no reliable 
characterization of the metallic or
insulating behavior of different phases and transition points can be
given. 

\subsection{Gaps to excited states}
\label{subsec:exandssgap}

In this section, we calculate excited states within the $S_z=0$ sector.
Due to the additional numerical difficulty of calculating excited
states in the same quantum number sector, we are restricted
to systems of $L=450$  lattice sites for $\Delta_{\text{E}}$ 
and $L=256$ sites for $\Delta_{\text{SE}}$.

In Fig.\ \ref{fig:exgap_all}(a), $\Delta_{\text{E}}$ as a function of $U$ is
presented for $\delta=20$ and 
various $L$.
For comparison, the spin gap for $L=300$ is also shown.
The exciton gap develops a local minimum around 
$U=21.38$, which, for increasing $L$, becomes sharper. 
Furthermore, the value at the minimum becomes smaller and seems 
to approach zero.
There is a cusp in $\Delta_{\text{E}}$ for all system
sizes shown here 
at a certain $U$ to the right of the minimum, and then a
smooth decay towards zero gap with further increasing $U$.
As illustrated for $L=300$, this corresponds to a level crossing with
the first triplet ($S=1$) excitation, which becomes the first
excited state for all larger $U$ values, i.e., 
$\Delta_{\text{E}} = \Delta_{\text{S}}$.
The data for $\delta=1$ and $\delta=4$ behave similarly,
but the increase to the right of the minimum (up to the cusp)
is substantially steeper as a function of $\delta$, so that it
approaches a jump.

In Fig.\ \ref{fig:exgap_all}(b), we display $\Delta_{\text{E}}$, the gap to the
second excited state $\Delta_{\text{SE}}$, and the spin gap
$\Delta_{\text{S}}$ (calculated using the ground state in the $S_z=1$ sector)
for $L=128$.
It can be seen that $\Delta_{\text{SE}} < \Delta_{\text{S}}$ 
for $U$-values to the left of the minimum in $\Delta_{\text{E}}$.
A similar behavior is found for
$\delta=4$ and $\delta=1$.
This means that 
there is more than one $S=0$ 
excitation below the lowest lying 
$S=1$ excitation, 
consistent with a scenario in which a continuum of 
$S=0$ excitations
becomes gapless at $U_{\text{c1}}$. 
This is the scenario predicted to occur at the first
quantum critical point in the
bosonization approach.\cite{Fabrizio}
Since system sizes for calculations of $\Delta_{\text{SE}}$ were limited to
$L=128$ ($L=256$ for some parameter values), we did not attempt to
systematically extrapolate $\Delta_{\text{SE}}$ to the thermodynamic limit.

We next 
discuss the
finite-size scaling 
for $\Delta_{\text{E}}$ to the left of the 
cusp.
For $U$ sufficiently far from the critical region (i.e., the minimum),
the finite-size corrections are small and the data can safely be
extrapolated to the thermodynamic limit using a quadratic polynomial 
in $1/L$, leading to a finite exciton gap. 
Close to the minimum, the scaling becomes more complicated.
At smaller system sizes, we find $\Delta_{\text{E}}=\Delta_{\text{S}}$
and the scaling is nonlinear.
However, at larger system sizes, there is a crossover
to linear scaling 
with $\Delta_{\text{E}}(L) \neq \Delta_{\text{S}}(L)$. 
The crossover length scale becomes larger 
as $U$ approaches the position of the minimum.
As a consequence, a reliable finite-size extrapolation in the critical 
region requires very large system sizes.

\begin{figure}[t]
\includegraphics[width=0.45 \textwidth]
%                {psfiles/compare_deltas/exzitongap_Deltamin_alldelta_obc.eps}
		{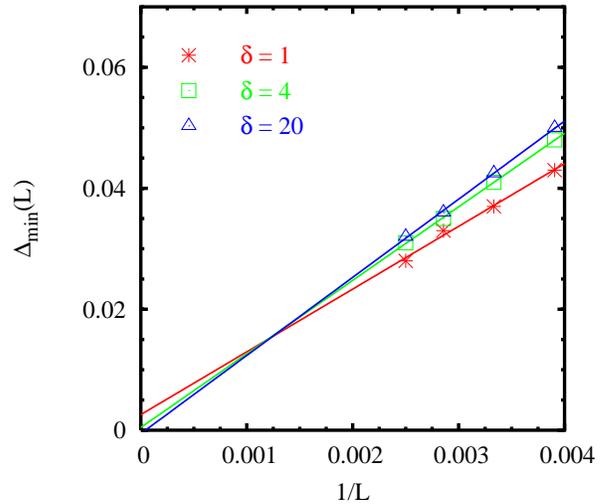}
\caption{Finite-size scaling analysis of the minimal value of the
  exciton gap $\Delta_{\text{E}}$.
  The solid lines are linear fits through the four system sizes shown,
  $L=256$, 300, 350, 400.
} 
\label{fig:exgap_extrapol}
\end{figure}

\begin{figure}[t]
\includegraphics[width=0.45 \textwidth]
%        {psfiles/delta20/excitongap/exzitongap_Uc_over_L_delta20_obc.eps}
		{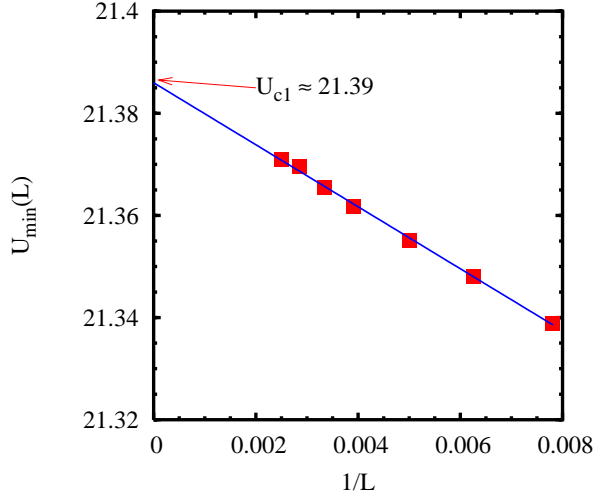}
\caption{Finite-size scaling analysis of the $U$-value at the minimum
  of the exciton gap $\Delta_{\text{E}}$ for $\delta=20$. 
  The solid line represents a linear least-squares extrapolation of
  the data yielding $U_{\text{c1}}  \approx 21.39$. 
} 
\label{fig:exgap_Uc}
\end{figure}

To investigate
the behavior as $L\to\infty$,
we interpolate $\Delta_{\text{E}}$ as a function of $U$ for fixed $L$ 
close to the minimum with
cubic splines. 
From the interpolation  we can read off the minimal
value of the gap $\Delta_{\rm  min}(L)$ and the position $U_{\rm min}(L)$
for the different system sizes.
Fig.\ \ref{fig:exgap_extrapol} shows the resulting 
$\Delta_{\rm min}(L)$ as a function of $1/L$ for $\delta=1$, $4$, and
$20$.
A linear extrapolation of the data
gives $\Delta_{\rm min}(L=\infty,\delta=1)=3\times10^{-3}$, 
$\Delta_{\rm min}(L=\infty,\delta=4)=5\times10^{-4}$, and  
$\Delta_{\rm min}(L=\infty,\delta=20) = -1\times10^{-4}$. 
Within the accuracy of our data and our extrapolation,
these minimal gaps 
can be considered to be  
zero.
In analogy with 
the atomic limit, we interpret
the vanishing of the exciton gap as defining a critical 
point.\cite{Sachdev} 
The critical coupling $U_{\text{c1}}$ can be determined from 
fitting $U_{\rm min}(L)$ to a linear function in $1/L$,
as shown for $\delta=20$ in
Fig.\ \ref{fig:exgap_Uc}.
The extrapolation 
is similar for the other $\delta$-values and we obtain
$U_{\text{c1}}(\delta=1) \approx 2.71$, $U_{\text{c1}}(\delta=4) \approx 5.61$, and
$U_{\text{c1}}(\delta=20) \approx 21.39$. 
As will be 
discussed in Sec.\ \ref{subsec:polsus}, the vanishing of the exciton gap
is accompanied by a diverging electric susceptibility. 

In Sec.\ \ref{subsec:bow}, we will present strong evidence in favor of a
spontaneously dimerized phase for $U_{\text{c1}}<U<U_{\text{c2}}$. 
Since the dimerized phase has an Ising-like symmetry, 
as $L \to \infty$ the ground state
in this phase is expected to be two-fold degenerate and the exciton
gap $\Delta_{\text{E}}$ is expected to vanish -  
at least if the thermodynamic limit is taken using PBC's.
At first glance this appears to be at odds with the increase of
$\Delta_{\text{E}}$ as a function of $U$ to the right of 
$U_{\text{c1}}$ (but before
the cusp is reached) as can be observed in Fig.\
\ref{fig:exgap_all}(a).
For finite systems, the OBC's
lift the degeneracy between the states with the 
two possible bond alternation patterns (strong, weak, strong, $\ldots$
and weak, strong, weak, $\ldots$), energetically favoring one of them 
which 
becomes the ground state.
We have calculated the bond
expectation values (see Sec.\ \ref{subsec:bow}) of the ground state and
the first excited state on systems of up to  $L=450$ (the largest size we were
able to reach) and find that the first excited state does not have the
opposite alternation pattern.
Instead, the alternation pattern is the same as in the ground state near
the ends, but reverses itself in the middle of the chain.
This change in the alternating BO parameter is evenly spread over the
chain so that it has a cosine-like form with two nodes.
It is difficult to perform finite-size extrapolation on
$\Delta_{\text{E}}$ in this region since there are few system sizes
and only a very limited range of $U$ available.
However, one might speculate that $\Delta_{\text{E}}$ will remain
finite as $L \to \infty$ due to the pinning of the BO parameter at
the ends.

Sufficiently far from $U_{\text{c1}}$, the data presented in Fig.\ 
\ref{fig:exgap_all}(a) suggest a 
linear closing of the exciton gap, which gets rounded off 
in the critical region for finite systems.
The larger $L$, the closer to $U_{\text{c1}}$ the deviation from linear
behavior sets in.
This suggests that
$\Delta_{\text{E}} \sim U_{\text{c1}} - U$ close 
to but below the first critical point.
It implies that the product of the critical exponents 
$z_1 \nu_1 = 1$ 
at the first critical point,\cite{Sachdev} 
where $z_1$ is the dynamical critical exponent and $\nu_1$ is the
exponent associated with the divergence of the correlation length.

Our finding of a vanishing exciton gap at the coupling $U_{\text{c1}}$ 
for OBC's is consistent with results obtained using PBC's and 
$L=4n$. 
For this case, a ground-state level crossing of two spin
singlets at $U=U_{\text{x}}(L,\delta)$ (implying a zero exciton gap) was 
found using exact diagonalization of small 
systems.\cite{Gidopoulos,Torio,Brune} 
A change of the site inversion symmetry at $U=U_x$ 
was also observed.
In Sec.\ \ref{sec:pbc}, we will argue
that $U_{\text{x}}(L \to \infty, \delta)$ 
coincides with $U_{\text{c1}}(\delta)$.\cite{Torio}
The presence of the ground-state level crossings might lead one to
speculate that discontinuous behavior will persist in the
thermodynamic limit, implying a first order phase transition
at $U_{\text{x}}(L = \infty, \delta)$.
However, we find no discontinuous behavior for systems with 
OBC's, either on finite systems or in the $L\to \infty$
extrapolations.
In order to agree with the results obtained for OBC's 
in the thermodynamic limit,
the discontinuous behavior for PBC's must 
become progressively 
smoothed out as 
$L\to \infty$.
\begin{figure*}[t]
\includegraphics[width=0.45 \textwidth]
%		{psfiles/delta20/spingap/spingap_all_delta20_obc.eps} 
		{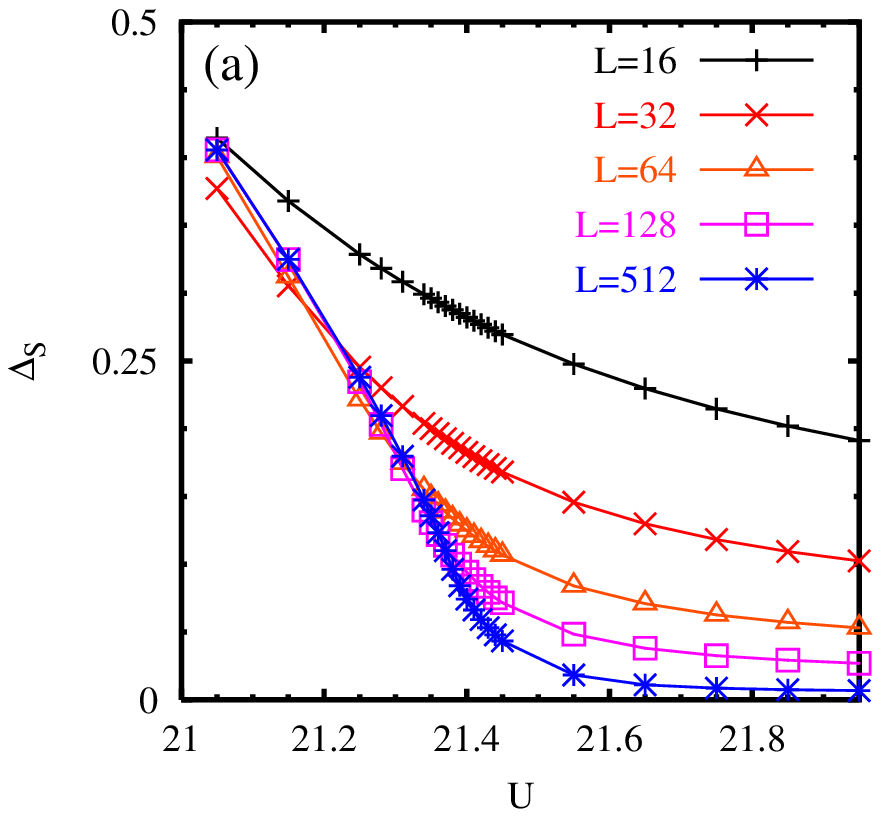}
\includegraphics[width=0.465 \textwidth]
%		{psfiles/delta20/spingap/spingap_scaling_delta20_obc.eps}
		{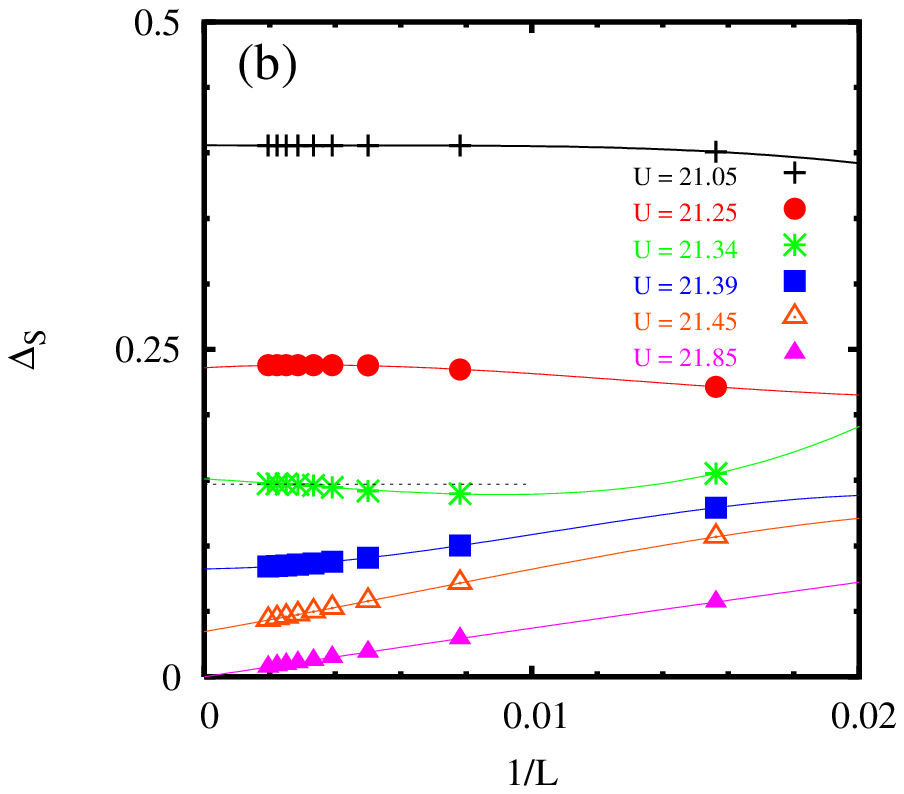}
\caption{(a) The spin gap $\Delta_{\text{S}}$ for finite systems 
   $L$ as a function of $U$ and (b) the
  finite-size scaling analysis for $\delta=20$ for chosen $U$-values. 
  The system sizes $L=64$, 128, 200, 256, 300, 350, 400, 450,
  and 512 are shown in (b) and are used for a least-squares fit to a
  third-order polynomial in $1/L$ (solid lines).
  The dashed line in (b) shows the value of $\Delta_{\text{S}}$ at the
  largest system size in order to illustrate the non-monotonic behavior.
} 
\label{fig:spingap_all}
\end{figure*}

\subsection{The spin gap $\Delta_{\text{S}}$}
\label{subsec:spingap}
The spin gap $\Delta_{\text{S}}$ 
is shown in 
Fig.\ \ref{fig:spingap_all} as a function of $U$ for $\delta=20$ and system
sizes between $L=16$ and 512.
In Fig.\ \ref{fig:spingap_all}(a), one can see that the spin gap
systematically scales towards zero above a certain $U$ value.
However, it is crucial that the finite-size scaling is carried out
carefully and systematically in order to determine the behavior in the
thermodynamic limit.
As can be seen in the scaling as a function of $1/L$ for
representative $U$ values in 
Fig.\ \ref{fig:spingap_all}(b), and as was pointed out in 
Ref.\ \onlinecite{Qin}, there is non-monotonic behavior as a function of
$1/L$ for $U < U_{\text{c1}}$.
In addition, the minimum of $\Delta_{\text{S}}$ as a function of $1/L$ 
shifts to larger system sizes as the critical region is approached.
This makes an extrapolation to the thermodynamic limit in the critical
region a difficult task which requires fairly large system sizes.
In order to carry out an accurate extrapolation, we fit to a cubic
polynomial in $1/L$. 

\begin{figure}[t]
\includegraphics[width=0.45 \textwidth]
%		{psfiles/delta20/gaps_TL/allgaps_TL_delta20_obc_multiplot.eps}
		{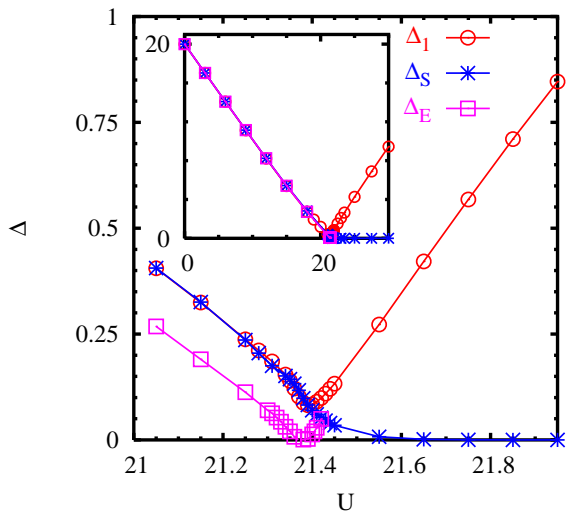}
\caption{The exciton gap $\Delta_{\text{E}}$, the spin gap $\Delta_{\text{S}}$, and the
  one-particle gap $\Delta_1$ for 
  $\delta=20$ after extrapolating to
  the thermodynamic limit $L \to \infty$. 
  The inset shows the result for a larger range of $U$. 
}
\label{fig:allgaps_TL_delta20}
\end{figure}

Fig.\ \ref{fig:allgaps_TL_delta20} shows the extrapolated
spin gap for $\delta=20$ presented together with the extrapolated
values for $\Delta_1$ and $\Delta_{\text{E}}$. 
All three gaps are
approximately equal for $U \ll U_{\text{c1}}$ (see the inset).
Close to the transition, as can be seen on the expanded scale in the
main plot, $\Delta_{\text{E}}$ goes to zero at $U_{\text{c1}}$, while
$\Delta_{\text{S}}$
and $\Delta_1$ stay finite and are (almost; see below) equal. 
For $U >  U_{\text{c1}}$, $\Delta_{\text{E}}$ increases until it
reaches the spin gap $\Delta_{\text{S}}$.
We find a region of $U > U_{\text{c1}}$ in which 
$\Delta_{\text{S}}(L = \infty)$ has a value that is clearly nonzero, well
above the accuracy of the data 
which is of the order of the symbol size. 
The behavior is similar for $\delta=4$ (not shown).
For even smaller values of $\delta$, $\Delta_{\text{S}}$ close to 
$U_{\text{c1}}$ becomes significantly smaller.
As a consequence, the region in which $\Delta_{\text{S}}$ is non-vanishing for
$U>U_{\text{c1}}$ is less pronounced at $\delta =1 $.
In this case, $\Delta_{\text{S}}$ at $U_{\text{c1}}$
is only a factor of six 
larger than the estimated accuracy of 
our data (this has to be compared to the factor of 20 for
$\delta=4$ and 40 for $\delta=20$) with a fast decrease for $U > U_{\text{c1}}$. 
We take the estimate of accuracy from comparison of DMRG calculations
for the one-particle gap of the usual 1D Hubbard model with Bethe ansatz
results.
We find that the difference is about 
$\left|\Delta_1^{\rm HM,DMRG}-\Delta_1^{\rm HM,exact}\right| = 0.003$ in the
worst case.
We nevertheless 
interpret this small spin gap to be finite for $\delta = 1$ and in a
small region of $U \geq U_{\text{c1}}$.
For $\delta$ 
substantially
smaller than $1$, it is impossible to
resolve a non-vanishing $\Delta_{\text{S}}$ at $U \geq U_{\text{c1}}$ 
using the DMRG.

The spin gap data in Fig.\ \ref{fig:allgaps_TL_delta20} indicate that
$\Delta_{\text{S}}$ 
goes to zero
very smoothly between $21.55$ and 
$21.8$ and remains zero from there on. 
We here define $U_{\text{c2}}$ 
as the coupling at which $\Delta_{\text{S}}$ goes to zero.
As we have argued in Sec.\ \ref{subsec:model}, the mapping onto a
Heisenberg model at strong coupling [Eq. (\ref{effectiveHeisenberg})]
suggests that the spin gap should vanish at sufficiently large $U$.
However, we cannot strictly speaking exclude that 
$U_{\text{c2}} = \infty$ from the spin gap data. 
We give further evidence in support of two transition points at finite
$U$ below.

Note that
the extrapolated (Fig.\ \ref{fig:allgaps_TL_delta20}) as well as the
large-$L$ data (Fig.\ \ref{fig:spingap_all})  
for $\Delta_{\text{S}}$ display an inflection point in the vicinity of $U_{\text{c1}}$. 
This might be an indication of a non-analyticity related to the  phase
transition at $U_{\text{c1}}$. 

\begin{figure*}[t]
\includegraphics[width=0.45 \textwidth]
%                {psfiles/delta20/1pgap/Ngap_all_delta20_obc.eps}
		{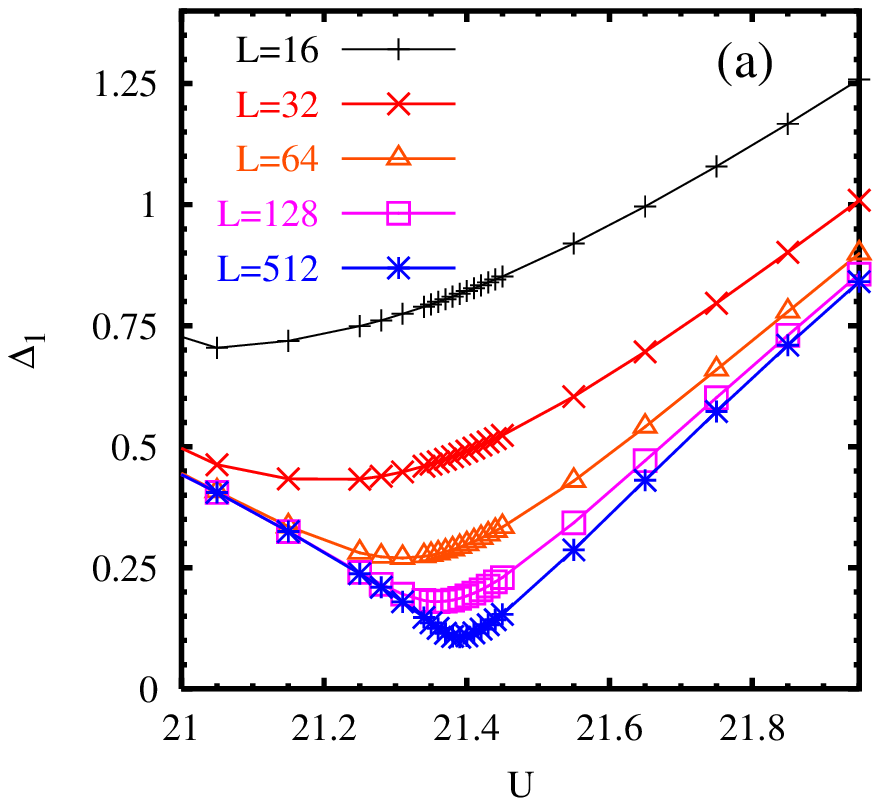}
\includegraphics[width=0.455 \textwidth]
%                {psfiles/delta20/1pgap/Ngap_scaling_delta20_obc.eps}
		{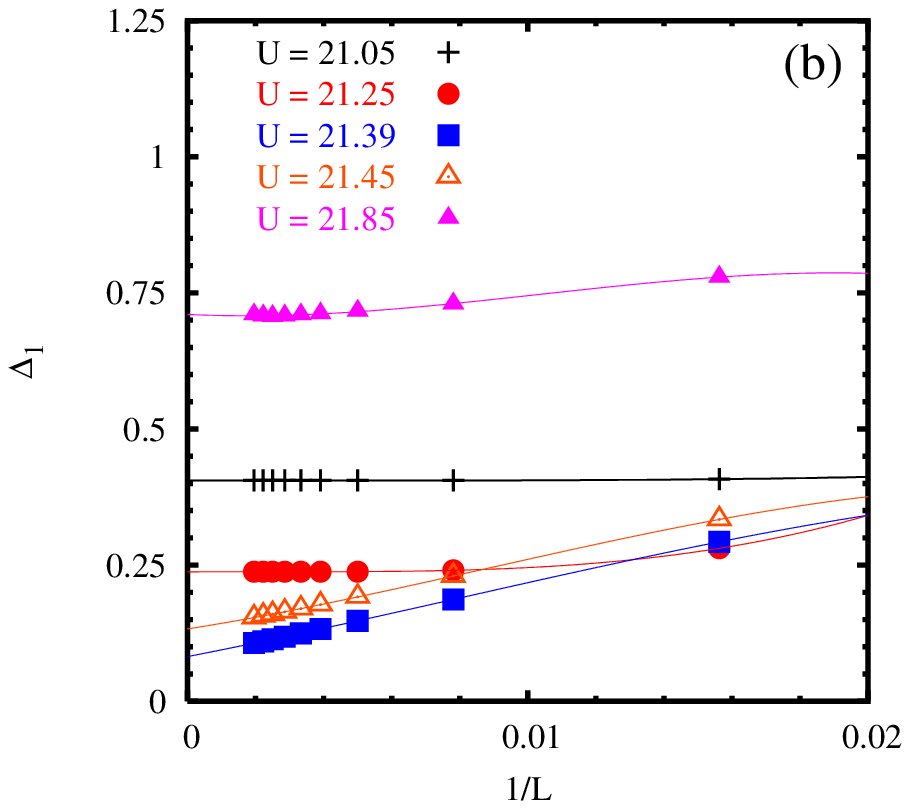}
\caption{The one-particle gap $\Delta_1$ for $\delta=20$. 
  (a) Results for finite systems with $L = 16$ through 512.
  (b) The finite-size scaling behavior for
  $L=64$, 128, 200, 256, 300, 350, 400, 450, 512. 
  The solid lines in (b) show least-squares fits to a third-order
  polynomial in $1/L$.
} 
\label{fig:1pgap_all}
\end{figure*}

\subsection{The one-particle gap $\Delta_1$}
\label{subsec:oneparticlegap}
In Fig.\ \ref{fig:1pgap_all}(a), $\Delta_1$ as a function of $U$ is shown
for $\delta=20$ and different $L$.
Away from the critical region 
(which is between $U \approx 21.15$ and $U \approx 22$), 
the finite-$L$ data rapidly approach the thermodynamic limit 
and accurate results for $\Delta_1(L=\infty)$ can easily be
obtained by fitting to a polynomial in $1/L$.
Close to $U_{\text{c1}}$, the data for large $L$ develop a minimum.
As $L$ increases, the position of the minimum
shifts to larger $U$-values.
The shape is quite rounded for the small system sizes, but becomes
sharper 
for the largest sizes.

In the critical region, the finite-size scaling is again delicate.
We examine $\Delta_1$ as a function of
$1/L$ for a number of $U$-values near $U_{\text{c1}}$ for $\delta=20$ 
in Fig.\ \ref{fig:1pgap_all}(b). 
The data sufficiently away from the minimum (on both sites) 
shows linear behavior in $1/L$ for smaller system sizes,
but then deviates from linear behavior and saturates at a finite value for
larger $L$-values.
This behavior is directly related to the $L$-dependence 
of the minimum of $\Delta_1$, which shifts to larger $U$ and becomes
sharper with increasing system size.
The scale on which a deviation from the linear behavior can be observed 
shifts to larger system sizes as $U$ approaches $U_{\text{c1}}$.
In order 
to perform the finite-size scaling 
analysis, we fit to cubic polynomials in $1/L$, as we did for the spin
gap.
We have carried out this procedure 
for $\delta=1$ and $4$ and find
that $\Delta_1(L,U)$ behaves similarly.

We have extracted the position and value at the minima by interpolating the
data for fixed $L$ 
with cubic splines and then extrapolating
to $L\to\infty$ with a fit to a quadratic polynomial.
We obtain 
$U_{\text{min}}(\delta=1) \approx 2.71$, 
$U_{\text{min}}(\delta=4) \approx 5.63$, and
$U_{\text{min}}(\delta=20) \approx 21.40$ for the positions and 
$\Delta_1(\delta=1,U_{\text{min}}) \approx 0.02$, 
$\Delta_1(\delta=4,U_{\text{min}}) \approx 0.05$, and
$\Delta_1(\delta=20,U_{\text{min}}) \approx 0.08$. 
The minimal values are finite to within the resolution of the
data and the extrapolation, although the values are small, especially
at small $\delta$.
Therefore, $\Delta_1$ is finite in the critical region and is
certainly larger than $\Delta_{\text{E}}$ which vanishes
at $U_{\text{c1}}$.
The positions of the minima are 
very
close to, but at a slightly larger
$U$-value than $U_{\text{c1}}$.
The largest difference $U_{\text{min}}(\delta)- U_{\text{c1}}(\delta)$ 
turns out to be $0.02$ (for $\delta=4$).
In Ref.\ \onlinecite{Qin}, 
calculation were carried out
for $\delta = 0.6$ (in our units), 
this difference was found to be $0.04$, and 
$\Delta_1(\delta,U_{\text{min}})$ was concluded to be zero. 
The authors  interpreted this as an
indication of a second transition point (in addition to
$U_{\text{c1}}$ which they
determined from the vanishing of the exciton gap).
While we have not carried out calculations at this value of $\delta$,
our results suggest that $\Delta_1(\delta=0.6,U_{\text{min}})$ is 
(perhaps unresolvably) small, but nonzero.
Therefore, we believe that $U_{\text{min}}$ is not associated with a 
second 
phase transition.
In fact, as we have seen in Sec.\ \ref{subsec:spingap}, the spin gap
goes to zero at a substantially higher value of $U$ than $U_{\text{min}}$, 
and we associate this value with $U_{\text{c2}}$.

Up to a small difference (see Fig.\ \ref{fig:allgaps_TL_delta20})
$\Delta_1(L=\infty)$ and $\Delta_{\text{S}}(L=\infty)$ are equal 
for $U < U_{\text{c1}}$.
In fact, the values are virtually identical for the largest few
system sizes and deviate only at smaller sizes.
We therefore believe that the difference in the extrapolated gaps
stems from differences in the fitting to the scaling function at
smaller system sizes and that $\Delta_1(L=\infty)=
\Delta_{\text{S}}(L=\infty)$ 
for $U < U_{\text{min}} \approx  U_{\text{c1}}$ 
is consistent with our results.
At this coupling,
$\Delta_1(L=\infty)$ starts to become larger than
$\Delta_{\text{S}}$ and as $U$ further increases, grows approximately linearly
in $U$ as one would expect in a Mott insulator.

To summarize the behavior of the 
finite-size extrapolated
gaps,
we find that for $U \ll U_{\text{c1}}$, 
$\Delta_{\text{E}}=\Delta_{\text{S}}=\Delta_1$ as in a
non-interacting band insulator.
As $U_{\text{c1}}$ is approached, the gaps to 
two (or more) $S=0$ excitations
drop below $\Delta_{\text{S}}$ and at least one of them 
goes to zero at $U_{\text{c1}}$.
The one-particle gap $\Delta_1$ reaches a finite minimum around 
$U_{\text{c1}}$ and then increases 
(linearly for large $U$), 
and the spin gap
$\Delta_{\text{S}}$ goes to zero 
smoothly
at $U_{\text{c2}} > U_{\text{c1}}$.
This smooth decay of the spin gap makes it difficult to
quantitatively estimate $U_{\text{c2}}$. 
Since the above behavior is similar for the widely different potential
strengths studied here, $\delta=1$, 4, and 20, we believe that it is
generic for {\it all} $\delta$.

\section{Ionicity}
\label{sec:cdw}

As argued in Sec.\ \ref{subsec:model}, the effective strong-coupling
model (\ref{effectiveHeisenberg}) predicts that the ionicity  
$\langle n_A -n_B \rangle \sim 1/U^3$ for large $U$.
For $t=0$, on the other hand, one expects a discontinuous jump from 
$\langle n_A -n_B \rangle = 2$ to $\langle n_A -n_B \rangle = 0$ at
the single transition point $U_{\text{c}}$.
Here we explore the behavior of $\langle n_A -n_B \rangle$ for all $U$
calculated within the DMRG.

\begin{figure}[t]
\includegraphics[width=0.45 \textwidth]
%                {psfiles/compare_deltas/naminusnb_log.eps}
		{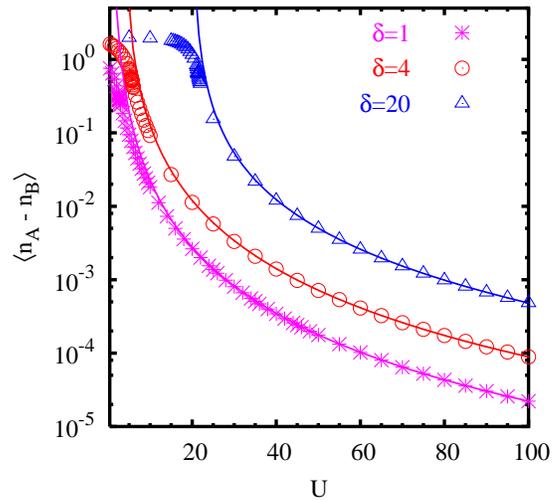}
\caption{The ionicity $\langle n_A - n_B \rangle$ for
  $\delta=1,4,20$. 
  The solid lines indicate analytical results from 
  Eq.\ (\ref{nAnBanalytical}) and the symbols numerical DMRG results for
  $L=32$ sites.
}
\label{fig:naminusnb}
\end{figure}

In Fig.\ \ref{fig:naminusnb} we compare Eq.\ (\ref{nAnBanalytical}) 
for $\delta=1$, 4, 20 and various $U$ to results 
obtained from DMRG with OBC's and $L=32$. 
By also considering larger system sizes (up to $L=512$)
and PBC's (up to $L=64$), we have verified that the $L=32$ results
shown are already quite close to the thermodynamic limit for 
$U \gg \delta$. On the scale of the figure the 
difference between $L=32$ and $L=\infty$ is negligible.  
For large $U$, the DMRG data agree quite well with the 
analytical prediction, Eq.\ (\ref{nAnBanalytical}). 
This gives a strong indication that the large-$U$ mapping of the IHM
onto an effective Heisenberg model\cite{Nagaosa} is 
applicable at large but finite $U$.
It is 
therefore 
tempting to conclude 
that $U_{\text{c2}} < \infty$.
One should nevertheless keep in mind that the excellent agreement of
the numerical data and the analytical prediction 
for the ionicity 
does not constitute a proof 
of this statement.
We will return to this issue.

\begin{figure}[t]
\includegraphics[width=0.45 \textwidth]
%        {psfiles/delta20/na_minus_nb/naminusnb_delta20_obc_multiplot.eps}
		{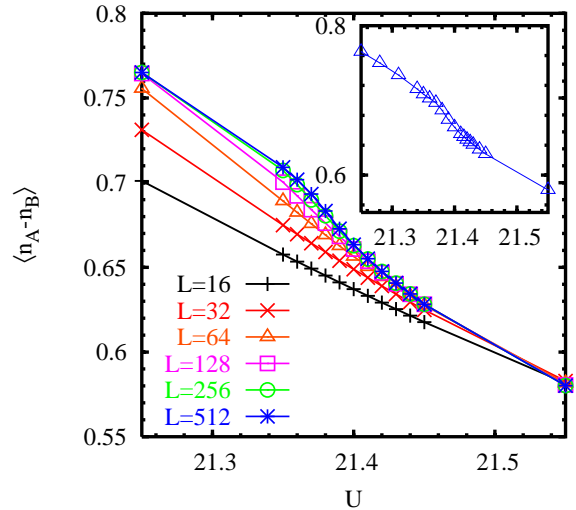}
\caption{The ionicity $\langle n_A - n_B \rangle$ for finite systems
  with $L=16$, 32, $\ldots\;$, 512 for $\delta=20$. 
  The inset shows the $L\to \infty$ extrapolated value.
} 
\label{fig:ionicity_critregion}
\end{figure}

The DMRG data for $\langle n_A - n_B \rangle$ for $L=32$ shown in 
Fig.\ \ref{fig:naminusnb} are continuous as a function of $U$ for all
$U$.
We examine $\langle n_A - n_B \rangle$ more carefully as a function of
system size in the vicinity of the first phase transition at $U_{\text{c1}}$ for
$\delta=20$ in Fig.\ \ref{fig:ionicity_critregion}.
The main plot shows DMRG data for various $L$ 
as a function of $U$ for $\delta=20$.
While the data are continuous as a function of $U$ for all sizes, there
is significant size dependence  
between $U=21.2$ and $21.5$, near the first critical point at
$U_{\text{c1}}$.
We have extrapolated the data to the thermodynamic limit using a
second order polynomial in $1/L$ and have checked that other
extrapolation schemes do not lead to  
significant differences in the extrapolated values.  
The $L=\infty$ extrapolated curve is shown in the inset.
While the curve is still continuous,
an inflection point can be observed close to $U_{\text{c1}}$.
This might be related to non-analytic behavior at $U_{\text{c1}}$.
We have found similar behavior of $\langle n_A - n_B \rangle$ for
$\delta=1$ and $4$.

The behavior of $\langle n_A -n_B \rangle$ for PBC's (and $L=4 n$),
which we have checked 
using the DMRG for up to $L=64$,
is quite different.
For finite $L$, the data display a jump discontinuity in the
critical region which decreases in size for increasing $L$.
The origin of this jump is the ground-state level crossing at 
$U_{\text{x}}(L,\delta)$.
Since we do not observe any discontinuity 
in the ionicity calculated for OBC's
for $\delta=1$, 4, 20 and up to $L=512$, and since the jump obtained for
PBC's becomes smaller with system size, we 
expect 
that the jump
vanishes in the thermodynamic limit 
and 
$\langle n_A -n_B \rangle$ becomes a continuous function.

\section{Order parameters and susceptibilities}
\label{sec:orderparams}
\subsection{The bond order parameter and susceptibility}
\label{subsec:bow}

The energy gaps have 
given us indications for two critical points. 
To study the nature of the intervening phase and the possibility of
dimerization in more detail, we calculate the BO parameter
\begin{eqnarray}
\left\langle B \right\rangle = \frac{1}{L-1} \sum_{j, \sigma} (-1)^j 
\left\langle c_{j+1 \sigma}^\dag c^{\phantom{\dag}}_{j \sigma} 
+ c_{j \sigma}^\dag c^{\phantom{\dag}}_{j+1 \sigma } \right\rangle \; .
\label{BOdef}
\end{eqnarray}
Since the OBC's break the symmetry between even and odd bonds,
$\left\langle B \right\rangle \neq 0$ for all finite systems.
Therefore, a spontaneous dimerization can be 
obtained
directly by
extrapolating  $\langle B\rangle$ to $L\to \infty$, i.e., without
adding a symmetry-breaking field explicitly.
One can form the corresponding BO susceptibility $\chi_{\rm BO}$ 
by adding a term
\begin{eqnarray}
H_{\rm dim} = \rho
 \sum_{j, \sigma} (-1)^j 
\left( c_{j+1 \sigma}^\dag c^{\phantom{\dag}}_{j \sigma} 
+ c_{j \sigma}^\dag c^{\phantom{\dag}}_{j+1 \sigma } \right) 
\label{HBOdef}
\end{eqnarray} 
to the Hamiltonian (\ref{Hamiltonian}) and taking
\begin{eqnarray} 
\chi_{\rm BO} = \left. \frac{\partial \langle B
\rangle(\rho)}{ \partial \rho} 
\right|_{\rho=0} \; . 
\label{chiBOdef}
\end{eqnarray} 
In practice, the derivative is discretized as 
$[ \langle B\rangle(\rho) - \langle B\rangle(-\rho)]/(2\rho)$ 
where $\rho$ is taken to be small enough so that the system remains in
the linear response regime.\cite{sizeoffieldamplitude}
Due to the additional symmetry breaking by the external dimerization
field $\rho$, the DMRG runs converge more rapidly than in the 
$\rho=0$ case, making it easier to reach larger system sizes. 
Thus we were able to calculate $\chi_{\rm BO}$ on
lattices of up to $L=768$ sites.
 
\begin{figure*}[t]
\includegraphics[width=0.45 \textwidth]
%                {psfiles/delta20/BO/bond_delta20_all_obc.eps}
		{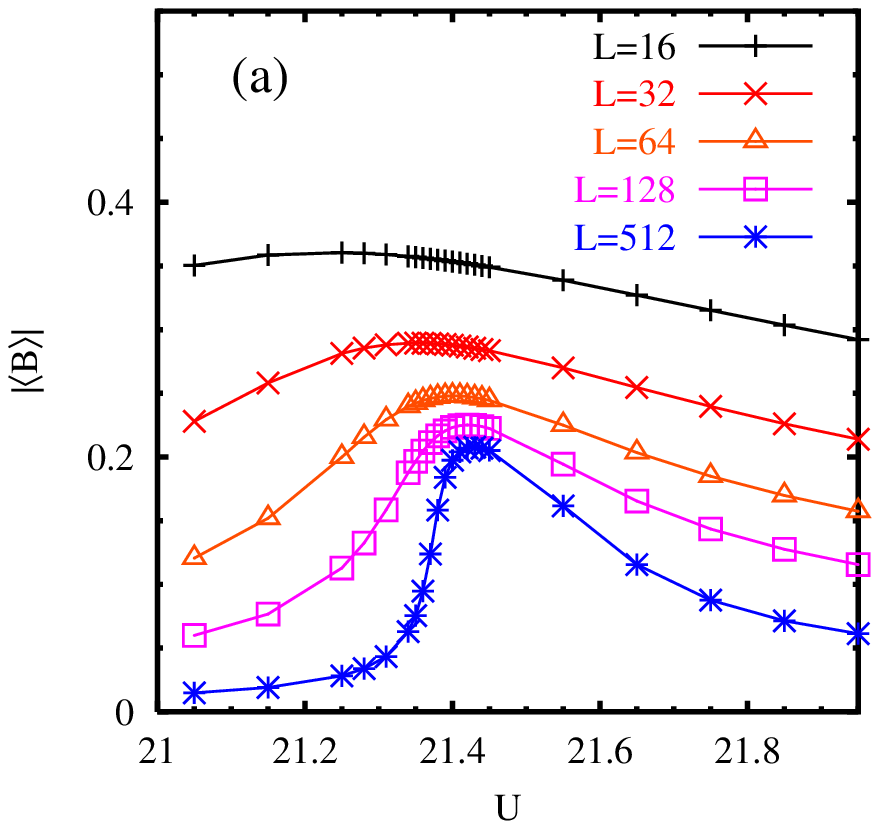}
\includegraphics[width=0.465 \textwidth]
%        {psfiles/delta20/BO/Lextrapol_BO_scaling_multiplot_delta20_obc.eps}
		{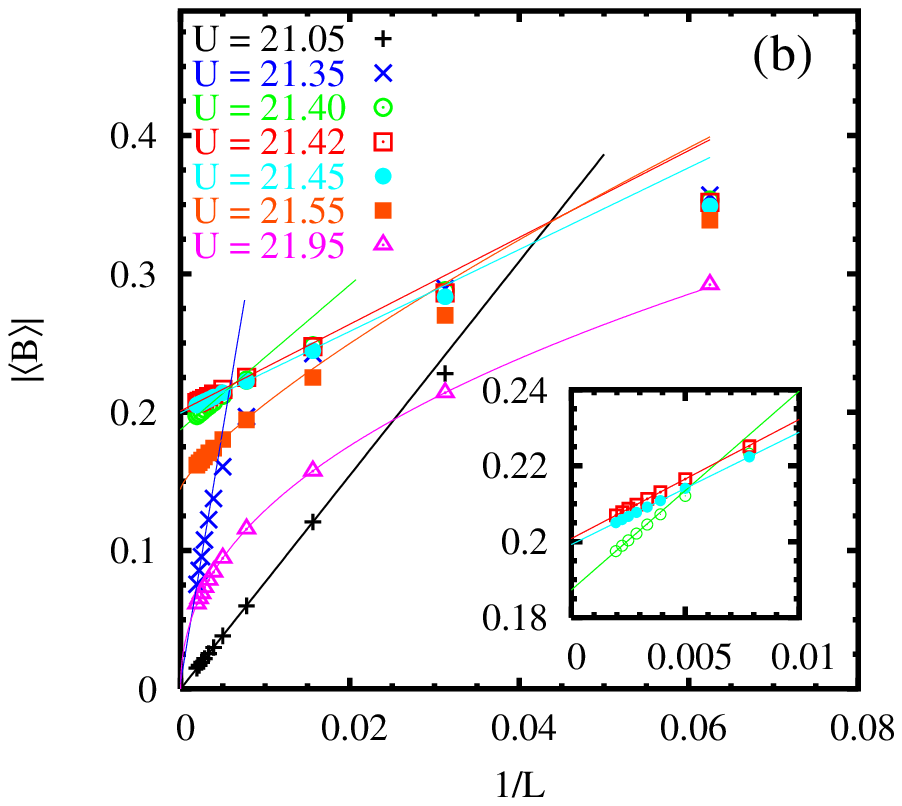}
\caption{(a) The bond order parameter $\langle B \rangle$ for $\delta=20$
  for finite systems as a function of $U$ for various system sizes.
  (b) The scaling of the data as a function of the inverse system size
  $1/L$.
  The solid lines are least-squares fits to the data as described in the text.
  The inset shows an expanded view of the scaling for $U$ values near
  the critical point.
} 
\label{BO_all}
\end{figure*}

Fig.\ \ref{BO_all}(a) shows $\left\langle B \right\rangle$ as a function
of $U$ for $\delta=20$ and 
different $L$.
The data develop a well-defined maximum near $U_{\text{c1}}$ for
large $L$. 
The width of the 
``peak''
for $L=512$ gives a first indication that there
is a region in which the dimerization is non-vanishing. 
Typical results for the finite-size scaling of 
$\left\langle B \right\rangle$
are presented in Fig.\ \ref{BO_all}(b). 
For $U \ll U_{\text{c1}}$, the data
extrapolate linearly to zero  in $1/L$. 
In the opposite limit, $U \gg U_{\text{c1}}$, we find 
$\left\langle B \right\rangle \sim 1/L^{\kappa}$ with 
$\kappa \approx 0.5 - 0.6$.
A similar slow decay of the BO parameter has also been found 
in the standard and extended Hubbard models at
half-filling.\cite{Eric}
The substantial finite-size corrections thus require very large
systems to distinguish between scaling to zero with 
a slow power-law and scaling to a finite $L \to \infty$ limit. 
Below, but close to $U_{\text{c1}}$, the data for small $L$ initially 
display power-law-like finite-size scaling with $\kappa < 1$, 
but for larger system size, one finds a crossover to a linear
scaling of the BO parameter (to zero) as $L \to \infty$.
There is also a crossover in the behavior for $U$-values 
near but above $U_{\text{c1}}$.
One again finds a crossover from a
power law with $\kappa <1$ for smaller system sizes 
to linear behavior that can be extrapolated to finite
values of $\langle B \rangle_{\infty}$ for larger system sizes.
The crossover length scale increases
as $U$ approaches $U_{\text{c1}}$ until it becomes larger than the largest
system size considered here.
This length scale $L_{\text{c}}$ can be used to estimate the
correlation length, which diverges at the first (continuous) critical
point.
We have been able to calculate $L_{\text{c}}$ for $U$-values on
both sides of $U_{\text{c1}}$ and find that it diverges 
approximately linearly in
$|U-U_{\text{c1}}|$.
This implies $\nu_1=1$ (see also below). 
Taking into account that 
$z_1 \nu_1=1$ as extracted from the linear closing of $\Delta_{\text{E}}$,
one finds $z_1=1$ for the dynamical critical exponent.

This diverging crossover length scale
makes it essential to treat 
system sizes that are significantly larger than the
scale $L_{\text{c}}$, even close to the critical point $U_{\text{c1}}$.
In order to obtain reliable results, we have calculated
$\langle B\rangle_L$ 
for a number of system sizes $L>200$. 
In carrying out the finite-size extrapolation, we fit to a linear form
for the largest system sizes if it is clear that $L_{\text{c}}$ has been
reached, as can be seen in 
the inset of Fig.\ \ref{BO_all}(b). 

\begin{figure}[t]
\includegraphics[width=0.45 \textwidth]
%		{psfiles/delta1/BO/BO_TL_delta1_obc.eps}
		{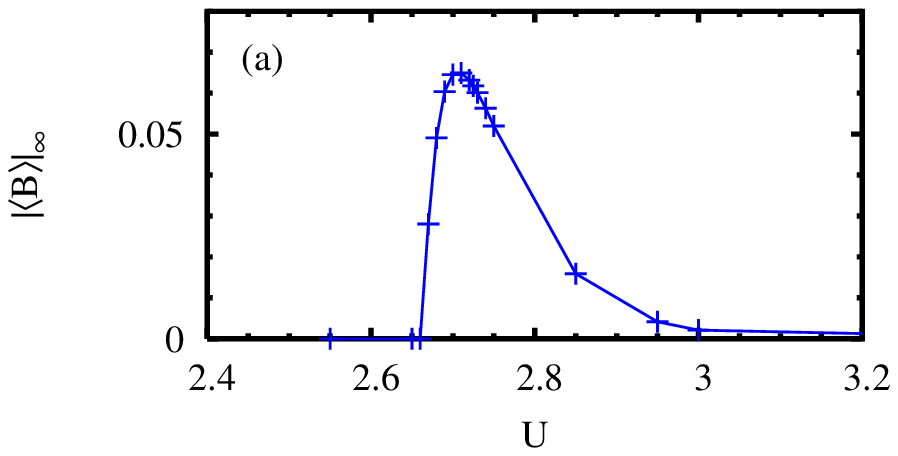}

\includegraphics[width=0.45 \textwidth]
%		{psfiles/delta4/BO/BO_TL_delta4_obc.eps}
		{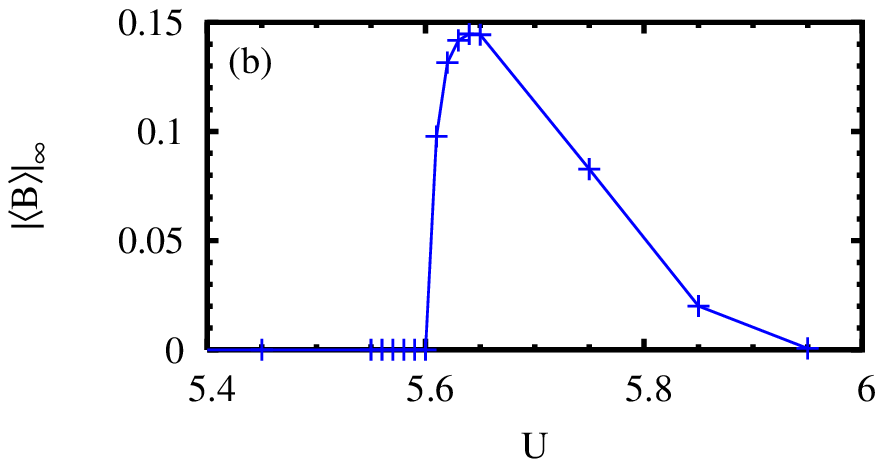}

\includegraphics[width=0.45 \textwidth]
%                {psfiles/delta20/BO/BO_TL_delta20_obc.eps}
		{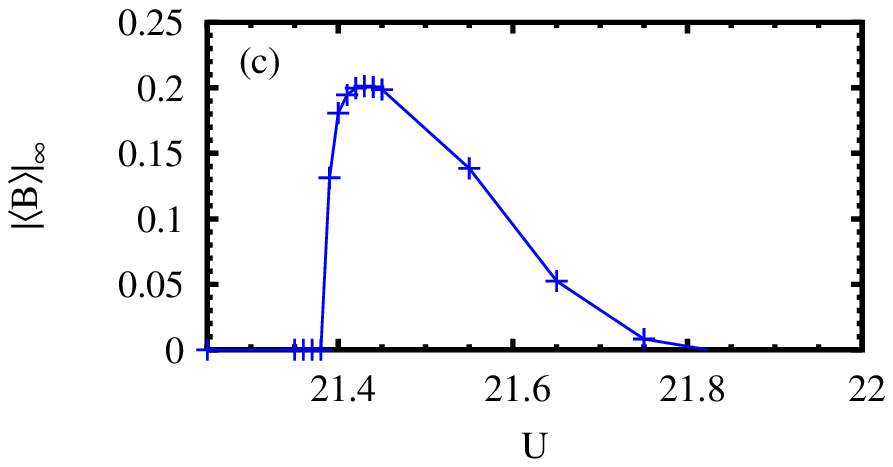}
\caption{The bond-order parameter $\langle B \rangle_{\infty}$ in the
  thermodynamic limit for (a) $\delta=1$, (b) $\delta=4$, and (c)
  $\delta=20$ plotted as a function of $U$ near the transition points.
} 
\label{fig:BOparam}
\end{figure}

In Fig.\ \ref{fig:BOparam}, the finite-size extrapolation
$\left\langle B \right\rangle_{\infty}$ is shown as a function of $U$ for 
$\delta=1$, 4, and 20. 
As can be seen, $\left\langle B \right\rangle_{\infty} = 0 $ to well
within the error of the extrapolation 
for $U < U_{\text{c1}}$. 
For $U > U_{\text{c1}}$, we find a region of width between 0.2 and 0.4 
(i.e., a factor of 5 to 10 larger than the extent of the dimerized
phase claimed to be found in Ref.\ \onlinecite{Qin}) in $U$ in which
$\left\langle B \right\rangle_{\infty}$ is distinctly finite.
The onset of finite $\left\langle B \right\rangle_{\infty}$ at
$U_{\text{c1}}$ is rather steep for all three values of $\delta$, but 
seems to be continuous.
This steep onset suggests a critical exponent of the order parameter 
that is substantially smaller than 1. 
Within bosonization the first critical 
point was predicted to be Ising-like with $\beta_1 =1/8$.\cite{Fabrizio}
The fall-off to zero as $U$ increases, on the other hand, 
is slow,
with a small or vanishing slope.
This behavior would be consistent with a second critical point 
at which
the critical exponent for the
order parameter is larger than one or at which 
a higher order phase
transition such as a Kosterlitz-Thouless transition takes 
place.\cite{Fabrizio}
As can be seen by comparing Fig.\ \ref{fig:BOparam}(a), (b), and (c),
the height of the maximum
increases with increasing $\delta$.  
For $\delta$ significantly 
smaller than 1, the BO parameter is so small that it cannot be
concluded to be finite within the numerical accuracy of the DMRG.
For the couplings at which the finite dimerization sets in 
we obtain  
$U_{\text{c1}}(\delta=20) \approx 21.39$ and 
$U_{\text{c1}}(\delta=4) \approx 5.61$,
which are in excellent agreement with the results obtained from the 
vanishing of $\Delta_{\text{E}}$. 
The value obtained for $\delta = 1$, 
$U_{\text{c1}}(\delta=1) \approx 2.67$, is also in reasonably good 
agreement with the results obtained from the
analysis of the gaps. 

While our data suggest that a critical coupling
$\tilde{U}_{\text{c2}}$, with $\left\langle B \right\rangle_{\infty}=0$ 
for $U > \tilde{U}_{\text{c2}}$, exists, no reliable quantitative estimate of
$\tilde{U}_{\text{c2}}$ can be given based on the DMRG data for the BO parameter. 
Due the close proximity of the two critical points, we 
were not able to obtain quantitative results for the 
critical exponents $\beta_1$ and $\beta_2$
at the critical points, either by a direct fit of the 
$L=\infty$ results or by a scaling plot of the finite-size data.
As 
discussed next, accurate exponents at 
$U_{\text{c1}}$ can be extracted from both the BO and the 
electric susceptibilities, and a more accurate estimate of 
$\tilde U_{\text{c2}}$ can be obtained from the BO susceptibility.

\begin{figure}[t]
\includegraphics[width=0.45 \textwidth]
%        {psfiles/delta20/BO/bond_manyssh_delta20_multiplot_allregimes.eps}
		{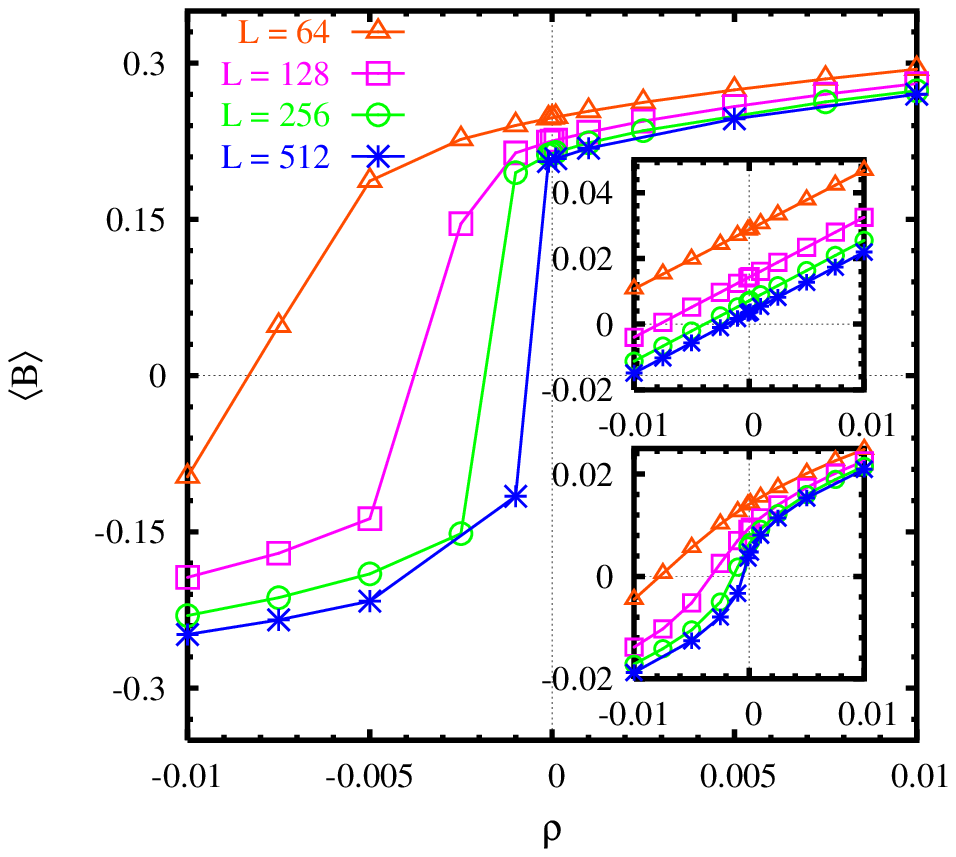}
\caption{
  The BO parameter $\langle B\rangle$ as a function of applied
  dimerization field $\rho$ for $\delta=20$ and $U=21.42$.
  The upper inset shows data for $U=19$ and the lower inset data for $U=50$.
}
\label{fig:BOparam_rho}
\end{figure}

In order to understand the behavior of the BO susceptibility, it is
useful to first examine the behavior of the BO parameter
$\langle B\rangle$ as a function of the applied dimerization field
$\rho$.
This quantity is shown in Fig.\ \ref{fig:BOparam_rho} for $\delta=20$,
three representative values of $U$, and different system sizes.
For $U = 19 < U_{\text{c1}}$, the system is in 
a phase with vanishing BO parameter,
and
the slope at $\rho=0$ remains finite 
for all system sizes,
corresponding to a finite susceptibility.
The value $U=21.42$ is in the intermediate regime where we have found
a finite BO parameter in the thermodynamic limit.
As can be seen in the main part of the figure, a jump in 
$\langle B\rangle(\rho)$ develops.
As the system size increases,
the absolute value of dimerization field at which the jump occurs 
becomes smaller.
This is the behavior expected in a dimerized phase 
in a system with OBC's.
Therefore, the jump in $\langle B\rangle(\rho)$ 
provides additional evidence in support of
an intermediate phase with finite dimerization.
For the approximate calculation of the susceptibility 
$\chi_{\rm BO} \approx [ \langle B\rangle(\rho) - \langle
B\rangle(-\rho)]/(2\rho)$, we have taken $\rho = 10^{-4}$ which is small 
enough to stay  to the right of the jump  
for all system sizes considered.
Finally, for $U=50 \gg U_{\text{c2}}$, 
$\langle B\rangle(\rho)$ goes to zero for $|\rho| \to 0$ and
increasing system size indicating a phase without spontaneous
dimerization. 
However, the slope at small $|\rho|$ becomes steeper with 
increasing system size, indicating a 
divergence of $\chi_{\text{BO}}$.

\begin{figure*}[t]
\includegraphics[width=0.45 \textwidth]
%                {psfiles/delta1/susc/bondsusc_all_delta1_obc.eps}
		{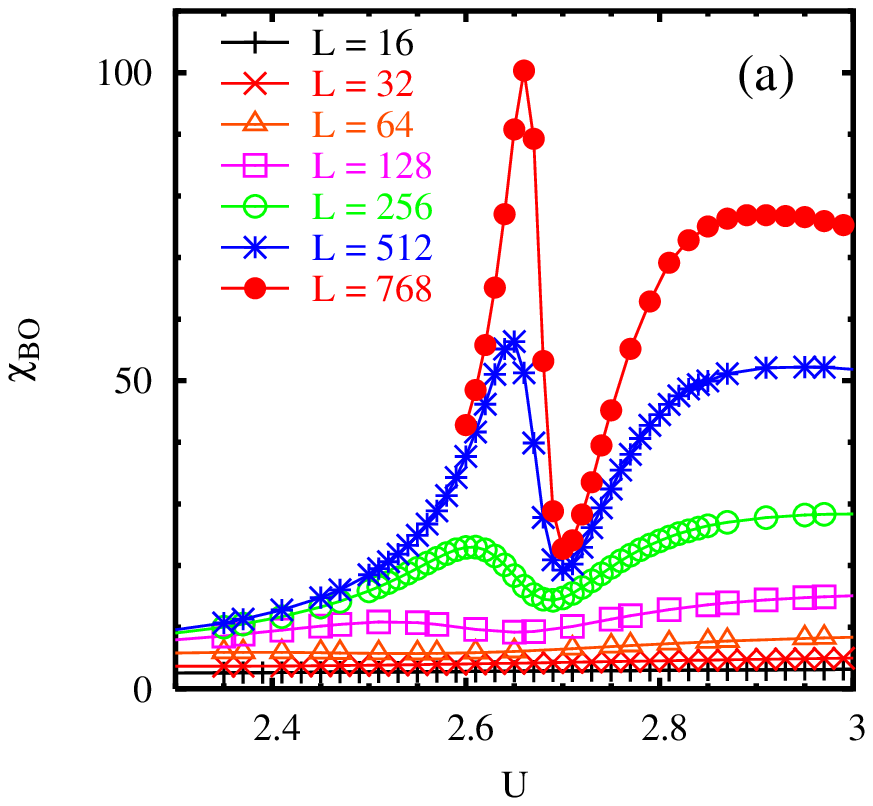}
\includegraphics[width=0.45 \textwidth]
%                {psfiles/delta4/susc/bondsusc_all_delta4_obc.eps}
		{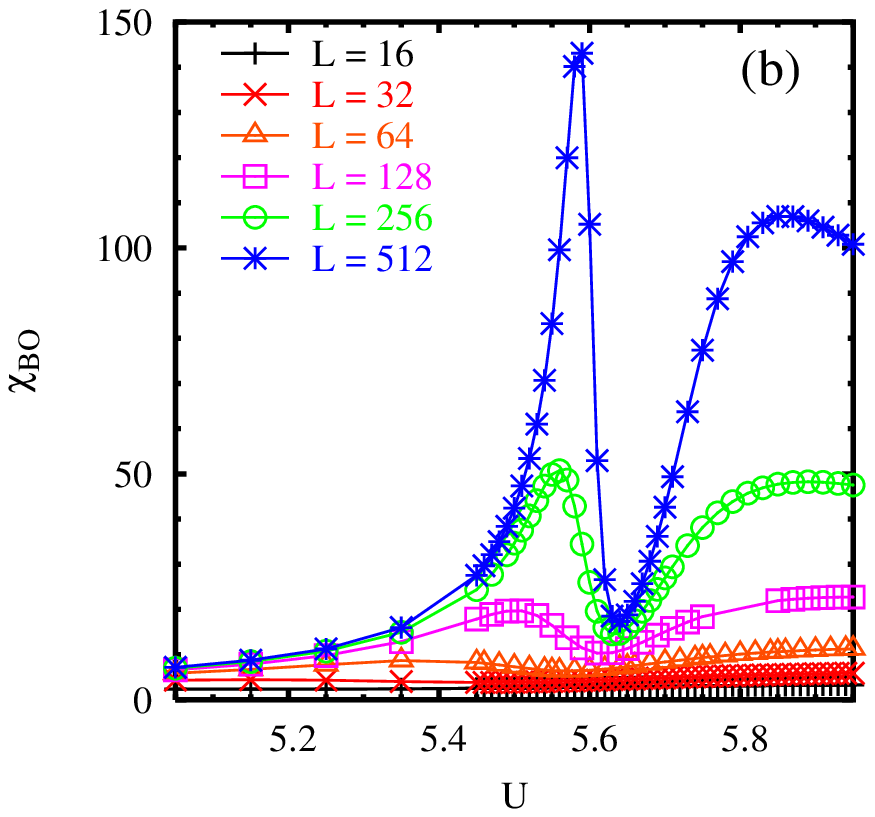}
\includegraphics[width=0.45 \textwidth]
%                {psfiles/delta20/susc/bondsusc_all_delta20_obc.eps}
		{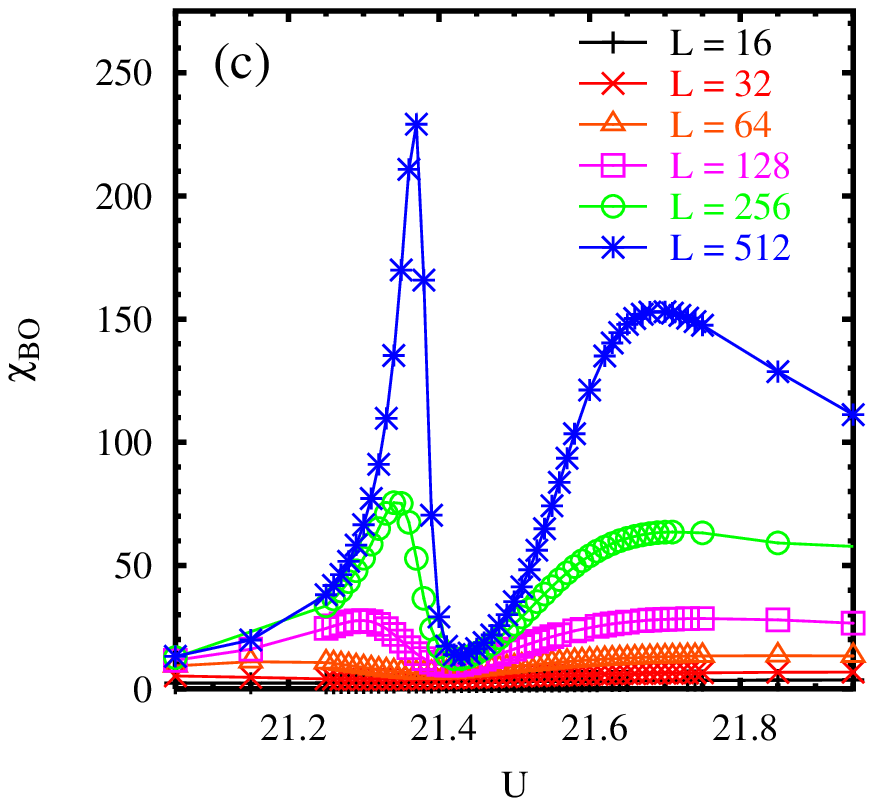}
\includegraphics[width=0.4625 \textwidth]
%                {psfiles/delta20/susc/scaling_bondsusc_delta20_Uc1_obc.eps}
		{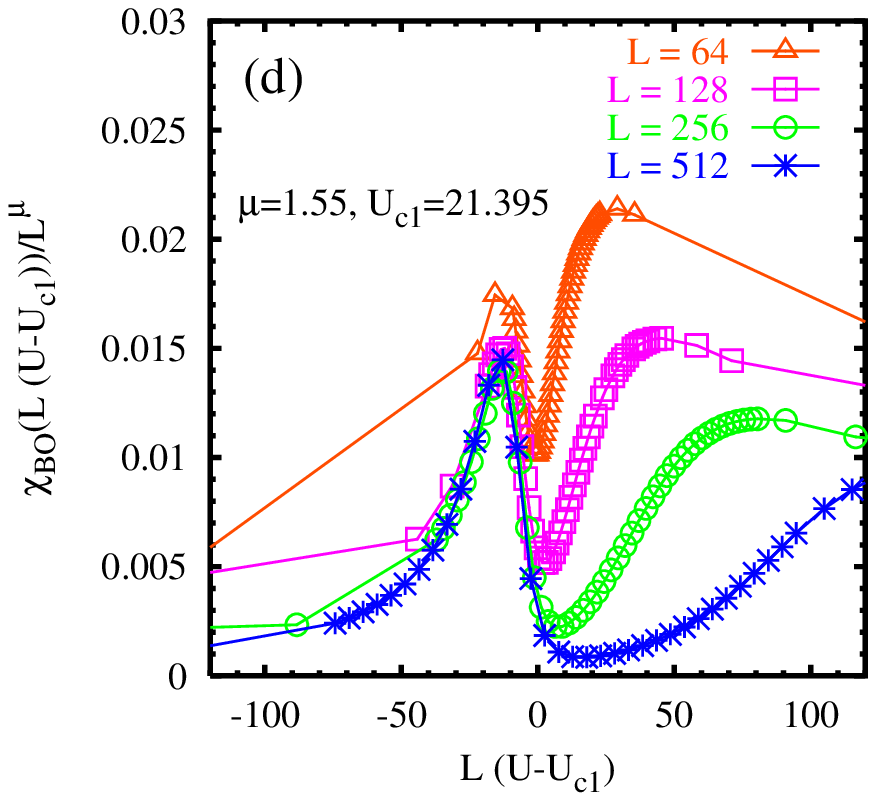}
\caption{The BO susceptibility $\chi_{\text{BO}}$ as a function of $U$
  for (a) $\delta=1$, (b) $\delta=4$, and (c) $\delta=20$ and
  different $L$.
  (d) A scaling analysis of the $\delta=20$ data from (c).
} 
\label{fig:BOsuscept}
\end{figure*}

\begin{figure}[t]
\includegraphics[width=0.45 \textwidth]
%            {psfiles/compare_deltas/extrapol_bondsusc_comparemodels_log.eps}
		{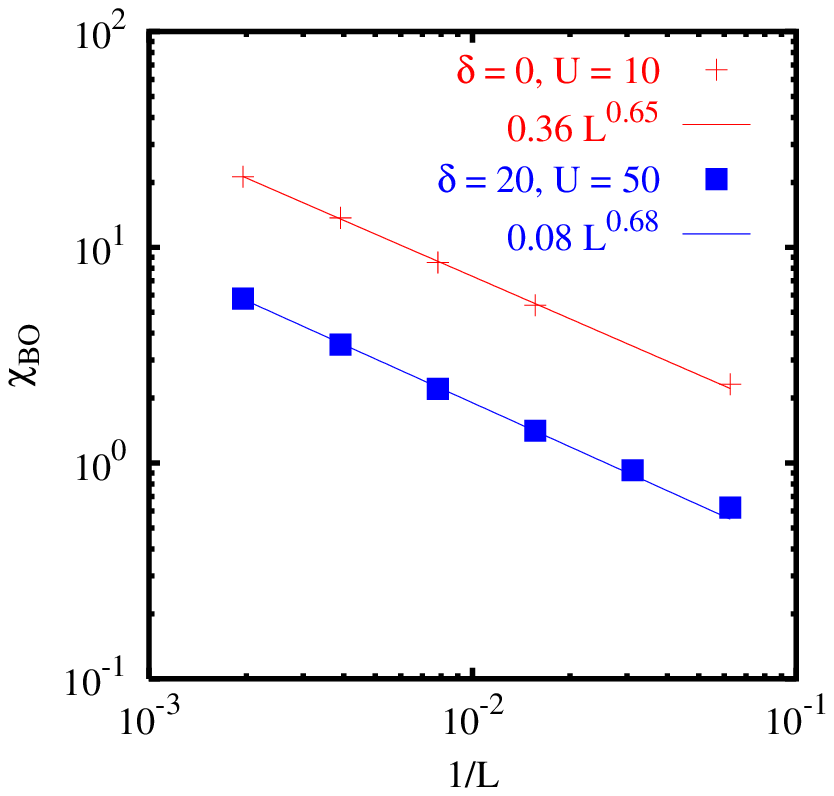}
\caption{The BO susceptibility $\chi_{\text{BO}}$ as a function of $1/L$ for
  the ordinary Hubbard model ($\delta=0$) and $U=10$ and the ionic
  Hubbard model for $\delta=20$ and $U=50$.
  DMRG data are indicated by the corresponding symbols and
  the solid curves represent a least-squares fit to the indicated forms.
}
\label{fig:bo_extrapol}
\end{figure}

In Fig.\ \ref{fig:BOsuscept}, the BO susceptibility as a function of $U$
is shown for $\delta=1$, 4, 20 and different $L$.
For all $\delta$ values, one observes  a two-peak structure that becomes
progressively more well-defined with increasing system size.
There is a narrow peak at a $U$-value that agrees well with $U_{\text{c1}}$
determined earlier
whose height grows rapidly with system size.
It signals the onset of spontaneous dimerization. 
For somewhat larger $U$ there is a minimum in $\chi_{\text{BO}}$, 
surrounded by narrow region in which 
its value seems to saturate with system
size.
For still larger $U$-values, a second, broad peak develops.
The position of this second maximum is roughly 
at $\tilde U_{\text{c2}}$, the $U$-value at which the BO parameter
vanishes. 
We argue that the second peak is related to the
second phase transition from the dimerized phase into an undimerized
phase. 
To the right of the second peak,
$\chi_{\text{BO}}$ does not seem to saturate for increasing system size,
implying that $\chi_{\text{BO}}$ is divergent for all $U\geq
U_{\text{c2}}$.
One can understand this divergent behavior by studying the 
BO susceptibility for the ordinary Hubbard model 
$\chi_{\text{BO}}^{\rm HM}$.
One finds that $\chi_{\text{BO}}^{\rm HM}$ is divergent for all $U>0$ because
the bond-bond correlation function 
is critical.\cite{Voit}
A finite-size extrapolation of $\chi_{\text{BO}}$ 
is shown in  Fig.\ \ref{fig:bo_extrapol} for large
$U$-values for both $\delta=0$ and $\delta=20$. 
We find a power law divergence, $\chi_{\text{BO}}(L) \sim L^{\zeta}$, with
$\zeta \approx 0.68$ 
for the ordinary Hubbard model and 
$\zeta \approx 0.65$ for the IHM.
These values are in good agreement, considering the accuracy of the
fit and additional finite-size effects.

Since $\chi_{\text{BO}}$ 
diverges for all $U$ to the right of the
second peak, it is 
difficult to accurately 
determine 
the critical coupling $ \tilde U_{\text{c2}}$. 
However, two different ways of estimating $\tilde U_{\text{c2}}(\delta)$
under- and overestimate its value.
In the first method, $\tilde U_{\text{c2}}$ is estimated as 
the lowest $U$-value
for which  $\chi_{\text{BO}}$ seems to diverge for increasing $L$ and the
available system sizes.
It is then still possible that there is a crossover 
above a length scale unreachable by us
and $\chi_{\text{BO}}$ scales to a finite value.

This 
tends to underestimate $ \tilde U_{\text{c2}}$.
In the second method, $\tilde U_{\text{c2}}$ is taken to be the position of
the second peak at fixed $L$, extrapolated to $L \to \infty$.  
Since the peak position decreases for increasing $L$,
this method tends to overestimate
$\tilde U_{\text{c2}}$. 
From these two procedures, we obtain the bounds 
$21.55<\tilde U_{\text{c2}}(\delta=20)<21.69$.
For the other values of $\delta$, it is very
difficult to accurately determine the lower bound with the data available. 
We therefore only give 
the upper bound 
$ \tilde U_{\text{c2}}(\delta=1)<2.95$ and 
$\tilde U_{\text{c2}}(\delta=4)<5.86$. 

It is generally believed that a
quantum critical point is accompanied by a vanishing characteristic 
energy scale.\cite{Sachdev} 
At $\tilde U_{\text{c2}}$ the most obvious 
candidate is $\Delta_{\text{S}}$, consistent with our numerical data 
(see Figs.\ \ref{fig:spingap_all} and \ref{fig:allgaps_TL_delta20})
and implying that $\tilde U_{\text{c2}} = U_{\text{c2}}$.
This is assumed in the following discussion.

Since the peak in 
$\chi_{\text{BO}}$
at $U_{\text{c1}}$ is well-defined and has a clear growth with
system size, it is 
reasonable
to perform a finite-size scaling
analysis.
We use a scaling ansatz of the form
\begin{eqnarray}
\chi(U,L) = L^{2-\eta} \tilde{\chi}(L/\xi) \; ,  
\label{scalinplot}
\end{eqnarray}  
with $\xi \sim | U-U_{\text{c}}|^{- \nu}$. 
As can be seen in Fig.\ \ref{fig:BOsuscept}(d), 
data for $\delta=20$ and system
sizes of $L=128$ and greater collapse onto one curve.
The best fit is obtained 
with $U_{\text{c1}} = 21.385$ and the critical exponents 
$\eta_{1}=0.45$ and $\nu_1=1$. 
The latter value is consistent with the value $\nu_1=1$ extracted from 
the divergence of the length scale discussed above.
Note that the value of $\eta_{1}$ is not in agreement with
the value expected in the two-dimensional Ising transition, 
$\eta= 1/4$.\cite{Fabrizio}
We have also applied the scaling ansatz for $\delta=1$ and $4$.
For decreasing $\delta$, the quality of the collapse of the data
for the available systems sizes becomes poorer and the extracted
exponents therefore become less reliable. 
The best fit is again obtained with $\nu_1=1$ for both $\delta$, 
$\eta_1(\delta=4) \approx 0.55$ and $\eta_1(\delta=1) \approx 0.65$.
For the critical coupling we obtain $U_{\text c1}(\delta=1)
\approx 2.7$ and $U_{\text c1}(\delta=4) \approx 5.6$, 
in excellent agreement with the values found by other means.

It is also possible to collapse the finite-size data onto 
one curve 
at the second transition point using the scaling ansatz
(\ref{scalinplot}). 
We find that the best results are obtained for 
$\xi \sim \exp(A/(U-U_{\text{c2}}^{\prime})^B)$, 
indicating that the divergence of
the susceptibility at $U_{\text{c2}}$ may indeed be exponential as expected
for a KT-like transition. 
However, fitting the limited amount of data available to this form
does not produce completely unambiguous results
for all fit parameters.

Therefore, we have not further attempted to obtain results
for $A$, $B$, $U_{\text{c2}}^\prime$, and $\eta_2$ with this method.

\subsection{The electric susceptibility and the density-density
correlation function}
\label{subsec:polsus}

In order to further investigate the 
physical properties of the different phases and transition points,
we calculate the 
electric polarization and susceptibility.\cite{Aebischer}
The polarization is given by 
\begin{eqnarray}
\left\langle P \right\rangle = \frac{1}{L} \sum_j x_j  
\left\langle n_{ j\uparrow} + n_{ j\downarrow} \right\rangle \; ,
\label{BOsuscdef}
\end{eqnarray}
where $x_j=j-L/2-1/2$ is the position along the chain, measured from
the center.
The polarization is the reponse due to a linear electrostatic potential
\begin{eqnarray}
H_{\rm el} = - E \sum_j  
x_j \left( n_{ j\uparrow} + n_{ j\downarrow}  \right) 
\label{HEdef}
\end{eqnarray} 
which is added to the Hamiltonian (\ref{Hamiltonian}). 
The electric susceptibility
\begin{eqnarray} 
\chi_{\text{el}} \; = \; \left. \frac{\partial \langle P\rangle(E) }{ 
\partial E} \right|_{E=0}
\label{chieldef}
\end{eqnarray}
is the susceptibility associated with this field. 

The electric susceptibility has been used to investigate the 
metal-insulator transition in the $t$-$t'$-Hubbard 
model.\cite{Aebischer} 
In this model, both a phase in which 
$\chi_{\text{el}}$ diverges as $L^2$ (a perfect metal) and 
a phase in which for increasing system size 
$\chi_{\text{el}}$ scales to a finite value (an insulator) 
were found when varying $U$ for fixed nearest-neighbor hopping $t$ and
next-nearest-neighbor hopping $t'$. 

In contrast to the ordinary Hubbard model, the polarization does not
always vanish at field $E=0$
in the IHM.
For $U=0$, $\delta>0$, one finds $\left\langle P \right\rangle = -1/2$.
This is due to the alternating ionic potential which induces a charge
displacement to the sites with lower potential energy.
Due to the OBC's, a chain with even length $L$ starts and ends with a
different potential, inducing a dipole moment.
This is a boundary effect. 
In the 
strong coupling limit, 
$U \gg \delta$, we find that  $\left\langle P \right\rangle \to 0$, as
expected.
The electric suceptibility $\chi_{\text{el}}$ can be calculated by
discretizing the derivative
as $[ \langle P\rangle(E) - \langle P\rangle(E=0)]/E$. 
The field $E$ must be taken to be small enough so that the system
remains in the linear response regime.\cite{sizeoffieldamplitude}
Note that it is necessary to subtract $\langle P\rangle(E=0)$
since it is nonzero in general.

\begin{figure*}[t]
\includegraphics[width=0.45 \textwidth]
%                {psfiles/delta20/susc/el_susc_delta20_obc.eps}
		{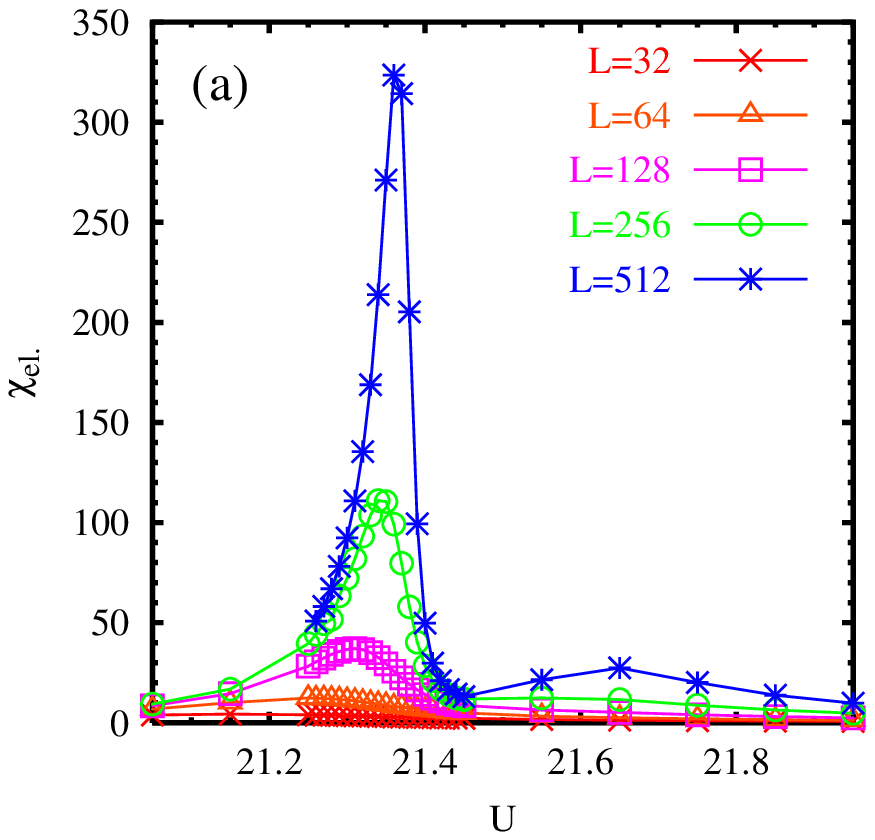}
\includegraphics[width=0.4775 \textwidth]
%                {psfiles/delta20/susc/scaling_el_susc_delta20_obc.eps}
		{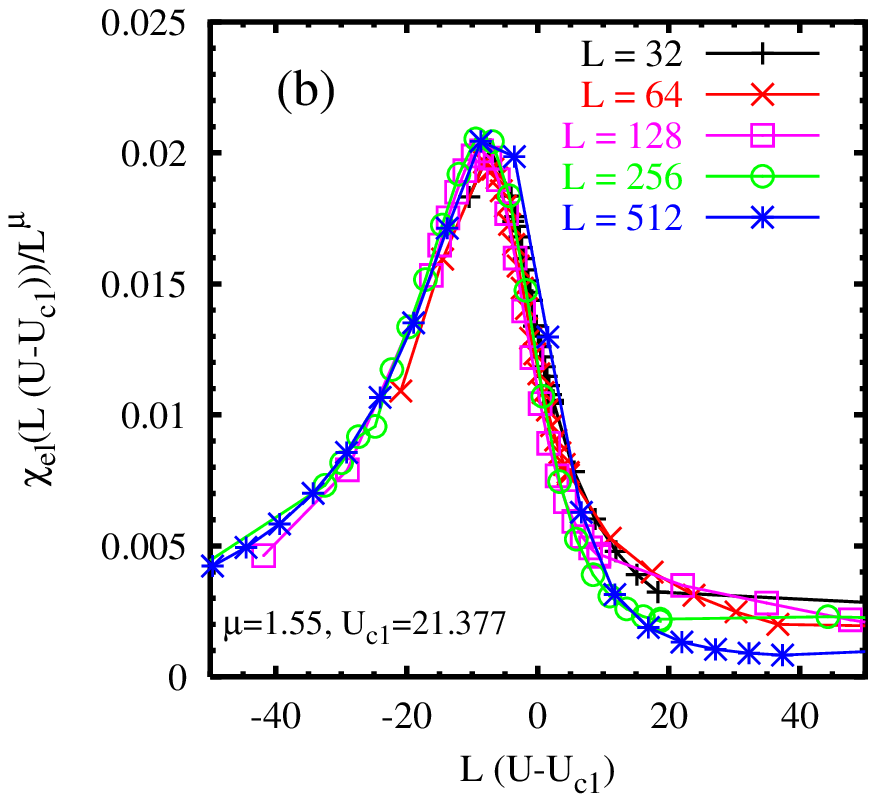}
\caption{
  (a) The electric susceptibility $\chi_{\text{el}}$ for
  $\delta=20$ plotted as a function of $U$.
  (b) A scaling analysis of the data of (a). 
} 
\label{fig:elect_suscept}
\end{figure*}

A plot of $\chi_{\text{el}}$ as a function of $U$ for various system
sizes is shown in Fig.\ \ref{fig:elect_suscept}(a)
for $\delta=20$. 
For $U \ll U_{\text{c1}}$ and increasing $L$, $\chi_{\text{el}}$ 
converges to a finite value, similar to the behavior in a 
non-interacting band insulator and in the correlated insulator phase of
the $t$-$t'$-Hubbard model.\cite{Aebischer}
The data clearly develop a maximum at $U_{\text{c1}}$ whose height 
increases markedly with system size, indicating a divergence at the first
critical point.
The finite-size scaling of this height
is consistent with a power-law increase, $L^{2 -\eta_{1}}$, with
$\eta_{1} \approx 0.46$. 
This increase is weaker than the $L^2$ 
divergence (which implies $\eta=0$) found in 
Ref.\ \onlinecite{Aebischer} and associated with a  
perfect metal. 
For $U$ slightly larger than $U_{\text{c1}}$, the data
again seem to saturate with system size. 
Assuming the scaling form of Eq.\ (\ref{scalinplot}), the data close
to $U_{\text{c1}}$ can be collapsed on a single curve as demonstrated in 
Fig.\ \ref{fig:elect_suscept}(b). 
The best fit is obtained for $\nu_1=1$
and $\eta_{1} \approx 0.45$. 
Both of these exponents are in excellent agreement with those found in
the scaling analysis
for $\chi_{\rm BO}$.
We have carried out a finite-size scaling analysis for $\delta=4$ and
$\delta=1$ and also find diverging peaks at $U_{\text{c1}}$, as well as
collapse of the data onto a single curve using the scaling form
(\ref{scalinplot}) with exponents $\eta_{ 1}(\delta=1)=0.52$,
$\eta_{1}(\delta=4)=0.45$, and $\nu_1=1$ (for both $\delta$).
The critical $U$-values obtained from this scaling procedure are
$U_{\text{c1}}(\delta=1) =2.68$, 
$U_{\text{c1}}(\delta=4) =5.59$, and $U_{\text{c1}}(\delta=20)=21.38$, 
which compare
well to the values for the critical coupling obtained from the gaps
and from the BO parameter and susceptibility.

The data for $\delta=20$ 
and $\delta=4$
for the largest system sizes, $L=256$ and
$L=512$, suggest that a second peak may develop around $U_{\text{c2}}$. 
In order to investigate the behavior of $\chi_{\text{el}}(L)$ more precisely
in this region,  
we fit a quadratic polynomial to $\langle P \rangle(E)$ 
through several data points and then take the derivative of this fit
function at $E=0$. 
This procedure should eliminate errors caused by a small 
linear response regime.
Results obtained from this procedure for $\delta=20$ indicate a weak 
divergence at $U=21.65$, corresponding to a $U$-value 
near $U_{\text{c2}}$. 
In addition, we find an even weaker divergence for {\it all} $U>21.65$. 
The larger the $U$-value, the smaller the coefficient of the
diverging part,
so that the divergence is very difficult to observe numerically deep in the
strong-coupling-phase.
One generally expects 
the divergence of $\chi_{\text{el}}$ to be connected to the closing
of a gap to excited states which possess at least some ``charge 
character'' (in the sense discussed below). 
At $U_{\text{c1}}$ the divergence is
accompanied by the closing of the exciton gap, 
leading to a 
consistent picture.   

\begin{figure}[t]
\includegraphics[width=0.45 \textwidth]
%                {psfiles/delta1/corr/256_denscorr_delta1_obc_largeU.eps}
		{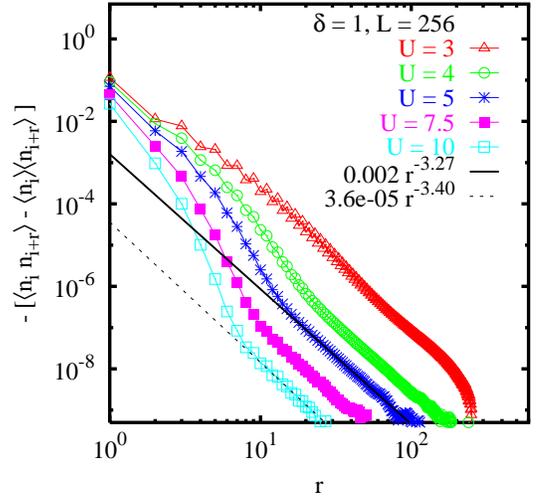}
\caption{
The negative of the 
density-density correlation function 
$ - \left [ \langle n_i n_{i+r} \rangle - \langle n_i \rangle \langle n_{i+r}
\rangle \right ]$ 
for $U>U_{\text{c2}}$. 
The indicated lines are least-squares fits over a range of
$r$ in which the behavior is linear on the log-log scale.
} 
\label{fig:denscorr}
\end{figure}

The situation is less clear for $U \geq U_{\text{c2}}$.
This issue can further be investigated by examining the behavior of the
density-density correlation function
\begin{equation}
C_{\rm den}(r) = \left\langle n_{i} n_{i+r} \right\rangle-
\left\langle n_i \right\rangle \left\langle n_{i+r} \right\rangle \;,
\label{denscorrDef}
\end{equation}
shown in Fig.\ \ref{fig:denscorr}
for $\delta=1$ and different $U > U_{\text{c2}}$. 
Here we have averaged over a number of $i$-values (typically six) for
each $r$ and have performed the calculation 
on an $L=256$ lattice.
For each value of $U$, it is evident that the 
correlation function
behaves linearly on the log-log scale above some value of $r$,
indicating that the dominant long-distance behavior is a power law.
(For $r$ close to the system size, finite-size effects from the open
boundaries also appear.)  
Note that the sign of the correlation function is negative for $r>0$, 
so that the negative is plotted.
A least-squares fit to the linear portion of the curve yields an
exponent of approximately
$3 - 3.5$ for all values of $U>U_{\text{c2}}$.
This behavior is markedly different from the behavior for 
$U < U_{\text{c1}}$, where we 
find a clear exponential decay as in a non-interacting band insulator, 
and from the behavior at $U_{\text{c1}}$, where we find a 
power law decay with an exponent of $\approx 2$. 
Note that if the decay were exponential
for $U>U_{\text{c2}}$, we would expect the
correlation length to change quickly with $U$, leading to a marked
variation in the slope.
We have ruled out finite-size effects as an origin of the power-law
tails as well as possible symmetry breaking due to the OBC 
by comparing calculations for $L=128$ and $L=256$
with OBC and $L=64$ with PBC,
which yield
identical values except for distances $r$ near 
the lattice size (or half the lattice size for PBC's).

We have performed calculations for $\delta=20$ and find
similar behavior.
The exponent of the power-law tails has a
comparable value 
to the ones given above, 
even at very large 
$U$-values such as $U=50$, where the
prefactor of the power-law part is $\approx 2 \times 10^{-6}$.
It therefore seems justified to conclude that this power-law decay is a
generic feature of the strong-coupling phase for all $\delta$. 

Our findings
for $\chi_{\text{el}}$ and $C_{\rm den}(r)$
are consistent with a scenario in which there is a continuum
of gapless excitations for $U>U_{\text{c2}}$, 
where matrix elements of charge operators such as 
the density
$n_j=n_{j\uparrow} +n_{j\downarrow}$, are 
nonvanishing for some of the states belonging to 
this continuum.
These are the states mentioned above which possess charge character. 
To further confirm this idea,
we have calculated matrix elements $\left \langle m \right| n_j \left| 0
\right\rangle $, where $\left| m\right\rangle$ denotes the $m$-th
excited state and $\left| 0\right\rangle$ the ground state, for up to 
$m=4$, $\delta=20$, $U > U_{\text{c2}}$, and $L=32$. 
We find that the third excited state is
the first $S=0$ state, both for the ordinary Hubbard model and the
IHM (the fourth state as well as the $m= 1$,2 states have $S=1$). 
For the ordinary 
Hubbard model, $\left\langle 3 \right| n_j \left| 0 \right\rangle$
vanishes for  all $j$ to within the accuracy of our data and this $S=0$ state 
can be classified as a spin excited state 
since its excitation energy is well below the charge gap.
In contrast, 
$\left<3 \right| n_j \left| 0 \right>$ is 
nonvanishing for the IHM and shows a 
non-trivial dependence on $j$
which has a wavelength of approximately the lattice size, implying
that the wave vector characterizing the excitation is near $q=0$.

As a consequence, this state contributes
to the dynamical charge structure factor in the IHM but not in the
ordinary Hubbard model. 
This shows that although several similarities
between the strong coupling phase of the IHM and the Hubbard model 
were found, low-lying excitations in both models are of quite different
nature.  As we have verified, the
energy of $ \left| 3 \right\rangle$ becomes smaller for increasing $U$, in
contrast to the behavior of the one-particle gap which increases
linearly with $U$. 
Due to the numerical effort necessary to target
such a large number of states, we were unable to perform
these calculations on larger lattices in order to carry out a finite-size
scaling analysis of the matrix elements. 

It is important to note that the power-law decay of $C_{\rm den}(r) $
and the divergence of $\chi_{\text{el}}$ for $U \geq U_{\text{c2}}$ do
not necessarily imply that the Drude weight is finite in this
parameter regime
or near $U_{\text{c1}}$, where
$\chi_{\text{el}}$ diverges roughly as $L^{1.5}$. 
Therefore, 
we refrain here
from classifying the dimerized phase, the $U \geq U_{\text{c2}}$ 
strong coupling phase, and the transition point $U_{\text{c1}}$ as
being metallic or insulating. 
For $U <U_{\text{c1}} $ all our results
are similar to those found in a non-interacting band insulator.
To further investigate the metallic and insulating behavior in
different parts of the phase diagram, it would
be necessary to calculate dynamical correlation functions using, 
e.g., the dynamical DMRG. 
Such an investigation 
would exceed
the scope of the current paper.

\section{Periodic boundary conditions}
\label{sec:pbc}

Up to this point, we have only presented DMRG results obtained for
systems with OBC's. 
Here we will argue that the results for energy gaps determined for
PBC's are consistent
with the ones discussed in Sec.\ \ref{sec:gaps}. 
We present further evidence
that $U_{\text{c2}}$ obtained from the closing of the spin gap 
for OBC's
and  the coupling constant $\tilde{U}_{\text{c2}}$ at which the BO
parameter vanishes coincide. 

Here we investigate the level crossing point of the ground state 
and the first excited state $U_{\text{x}}$, as well as the crossing of the 
first and second excited states $U_{\text{xx}}$. 
In Ref.\ \onlinecite{Torio} (and
further references therein), the crossing points of 
ground and 
excited
states of finite systems are associated with phase transition points.
In particular, the crossing of the ground state and the first excited
state
with opposite site-inversion parity were shown to correspond to a
jump in the charge Berry phase.
The crossing of the first and second excited states which are spin
singlets and spin triplets with opposite site-inversion symmetry and
zero total momentum was argued to be associated with a jump in the spin Berry
phase and to characterize a second transition.
While the direct calculation of the Berry phases is
beyond the scope of this paper,
it is possible for us to
analyze the finite-size behavior of the level crossings $U_{\text{x}}$ 
and $U_{\text{xx}}$.
We therefore must calculate
the energies of the ground state and the first two excited states 
simultaneously. 
Sufficiently accurate DMRG results for these energies can only be obtained 
for system sizes of up to $L=64$, small compared to the ones studied 
for OBC's, but nevertheless much larger than the ones that can be
reached with exact diagonalization.\cite{Torio,Gidopoulos,Pati,Brune} 
We show results for PBC's for systems with $L=12$, 16, 24, 32, and
64, i.e., system sizes with $4n$ sites,
so that the site-inversion symmetry of the ground state is guaranteed
to change sign with $U$,
as discussed in Sec.\ \ref{subsec:model}. 
We find non-monotonic behavior of the level crossing points as a
function of system size at a scale beyond the system sizes which can be 
investigated using exact diagonalization.

\begin{figure}[t]
\includegraphics[width=0.405 \textwidth]
%                {psfiles/delta1/pbc/aligia_delta1_multiplot.eps}
		{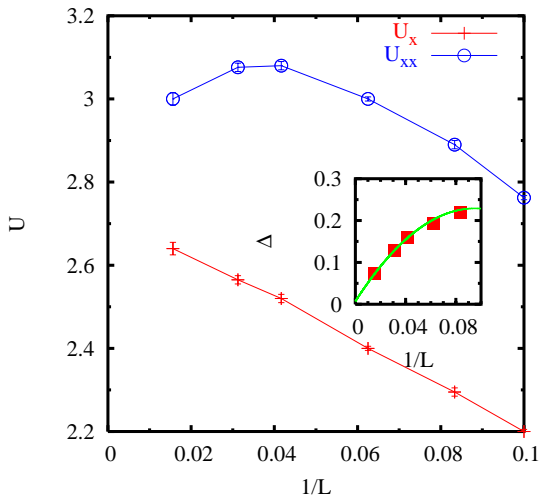}
\caption{Crossing points of excited states for PBC's for $\delta = 1$
  as a function of inverse system size.
  The inset shows the extrapolation of 
  $\Delta_{\text{S}}(U_{\text{xx}})$ as a function of inverse system size. 
} 
\label{fig:periodic}
\end{figure}

The finite-size scaling of $U_{\text{x}}$ and $U_{\text{xx}}$ for
$\delta=1$ 
is shown in Fig.\ \ref{fig:periodic}.
The error bars result from the uncertainty in
determining the closing or crossing points 
as well as from the poorer convergence of the DMRG algorithm for
PBC's. 
Nevertheless, for $\delta=1$ they only are of the order of the
symbol size 
or smaller.
For 
$\delta=4$ and $\delta=20$,
the convergence for $L=64$ is 
poorer around $U_{\text{x}}$ and
$U_{\text{xx}}$, but 
we obtain qualitatively similar behavior up to the larger error bars.

For all $\delta$ studied, 
the finite-size 
extrapolation of $U_{\text{x}}$ leads to critical couplings in agreement 
with the ones given in Sec.\ \ref{subsec:spingap}, up to the smaller
numerical accuracy available with PBC's.
Using a quadratic polynomial for the extrapolation, 
we find $U_{\text{x}}(\delta=1) = 2.71$, $U_{\text{x}}(\delta=4)=5.63$ 
and $U_{\text{x}}(\delta=20)=21.42$.
(Due to complicated finite-size effects, we only use the data for 
$L \leq 32$ for $\delta=4$ and $\delta=20$.) 
The angle of crossing of $E_0$ and $E_1$ decreases with increasing
system size.
This is consistent with a continuous critical behavior at $U_{\text{c1}}$ in
the thermodynamic limit.

The non-monotonic behavior of $U_{\text{xx}}$, as seen in 
Fig.\ \ref{fig:periodic}, makes an $L \to \infty$ extrapolation difficult. 
In fact, an extrapolation using the system sizes available to exact
diagonalization would give a $U_{\text{c2}}$ which is substantially larger
than if the two largest system sizes were included.
This could explain the discrepancy in the size of 
the region between the two critical points found here and obtained in 
Ref.\ \onlinecite{Torio}.
By extrapolating the finite-size data using a quadratic polynomial in
$1/L$, we obtain
$U_{\text{xx}}(L = \infty, \delta=1)=2.84$,  
$U_{\text{xx}}(L = \infty, \delta=4)=5.97$, and 
$U_{\text{xx}}(L = \infty, \delta=20)=21.75$.
The values for $U_{\text{xx}}$ and for $U_{\text{c2}}$ obtained from the
BO susceptibility
are in fairly good agreement.
The differences indicate
that even larger system sizes are needed to perform an accurate
finite-size extrapolation for PBC's.

Since the transition $U_{\text{c2}}$ is associated with the closing of the
spin gap, the gap to the excitations at $U_{\text{xx}}$ should scale to zero
with system size.
The inset of Fig.\  \ref{fig:periodic} shows that 
$\Delta_{\text{E}}(U_{\text{xx}})=\Delta_{\text{SE}}(U_{\text{xx}})$ 
indeed closes
in the thermodynamic
limit. 
Since one of the two states that are degenerate at $U_{\text{xx}}$ 
is a spin triplet, this implies a vanishing spin gap. 
We thus obtain 
further (indirect) evidence that the couplings at which the 
BO susceptibility diverges 
and the spin gap closes coincide, consistent with a 
two-critical-point scenario. 
In particular, the angle of the crossing
of the first and second excited state also decreases with increasing $L$, 
consistent with a 
continuous transition
at $U_{\text{c2}}$. 

\section{Summary}
\label{sec:sum}

In this paper, we have presented density-matrix
renormalization group results that elucidate the nature of the
quantum critical behavior found in  
the half-filled ionic Hubbard model.
By carrying out extensive and precise numerical calculations and by
carefully choosing the quantities used to probe the behavior, 
we have been able to investigate the structure of the transition more
accurately than in previous work. 
This has allowed us to
resolve a number of outstanding uncertainties and ambiguities.
We have worked at three different strengths of the
alternating potential $\delta$ 
covering a significant part of the parameter space
and find the same qualitative behavior 
for all three $\delta$-values.
In particular, we have carried out extensive finite-size scaling analyses of
three different kinds of gaps: 
the exciton gap, 
the spin gap, and the one-particle gap.
We find that 
for fixed $\delta$ and
in the thermodynamic limit, the exciton gap goes to zero
as a function of $U$ at a first critical point 
$U_{\text{c1}}$, the spin gap
goes to zero at a distinct second critical point 
$U_{\text{c2}} > U_{\text{c1}}$ and
is clearly nonzero at $U_{\text{c1}}$.
The one-particle gap (the two-particle gap behaves similarly) 
reaches a minimum 
close to $U_{\text{c1}}$, 
but never goes to zero and
never becomes smaller than the spin gap.

Due to the explicitly broken one-site translational symmetry,
the ionicity is finite for all finite $U$. 
For $U \gg \delta$
the ionicity found numerically agrees very well with the one 
obtained
analytically from the strong coupling mapping of the ionic Hubbard
model onto an effective Heisenberg model. 

We have also studied the bond-order parameter, the order parameter
associated with 
dimerization, as well as the associated bond-order
susceptibility.
The result of the
delicate finite-size extrapolation indicates that
there is a finite 
bond-order parameter in the intermediate region between $U_{\text{c1}}$ and
$U_{\text{c2}}$.
There is a divergence in the bond-order susceptibility at both
$U_{\text{c1}}$ and at $U_{\text{c2}}$, as one would expect from two continuous
quantum phase transitions.
However, the bond-order susceptibility diverges in the entire
strong coupling 
phase $U \geq U_{\text{c2}}$, albeit more weakly than at $U_{\text{c1}}$.
We have pointed out that this is 
in accordance with the behavior found in the strong coupling phase of
the ordinary Hubbard model.

We find that the electric susceptibility 
is finite for $U <U_{\text{c1}} $
but diverges 
roughly as $L^{1.5}$ at
$U_{\text{c1}}$.
This divergence is weaker than the one found for non-interacting
electrons 
(with $\delta=0$) 
and in the metallic phase of the 
$t$-$t'$-Hubbard model.\cite{Aebischer}  
A finite-size scaling analysis of both the bond-order susceptibility
and the electric susceptibility yield the same critical exponents at
$U_{\text{c1}}$.
However, the value, $\eta_1 \approx 0.45$,
is not
consistent with the critical exponents of the classical
two-dimensional Ising model.\cite{Fabrizio}

The electric susceptibility also seems to diverge, albeit quite
weakly, for $U \ge U_{\text{c2}}$.
Correspondingly, the density-density correlation function
has a long-distance
decay which is of power-law form, but with a small prefactor which
becomes smaller with increasing $U$, and a relatively large exponent
of approximately $3-3.5$.
We speculate that this behavior is
related to mixed spin and charge character of excitations 
present in the  strong coupling phase of the ionic
Hubbard model, 
in contrast to 
the ordinary Hubbard model.

We point out that the divergence of the electric susceptibility
at $U_{\text{c1}}$ and for $U \geq U_{\text{c2}}$
does not necessarily imply a finite Drude weight. 
Based on our results 
for 
various energy gaps and the electric susceptibility,  we
therefore cannot 
unambiguously classify all different phases and transition points 
as being metallic or insulating. 

Finally, we have presented DMRG results for the position of the
crossing of the ground state and the first excited $U_{\text{x}}$ and the
crossing of the first two excited states $U_{\text{xx}}$ 
on systems with periodic boundary conditions on up to 64 sites.
The finite-size extrapolation of $U_{\text{x}}$ gives
$U_{\text{c1}}$.
Due to the loss of accuracy, it is somewhat less clear 
that the finite-size extrapolation of $U_{\text{xx}}$ corresponds to 
$U_{\text{c2}}$.

\section*{Acknowledgements}
We thank 
A.\ Aligia, M. Arikawa, F.\ Assaad, D.\ Baeriswyl, P.\ v.\ Dongen, 
G.\ Japaridze, E.\ Jeckelmann, A.\ Kampf, A.\ Muramatsu, 
A.\ Nersesyan, B.\ Normand, M.\ Rigol, and M.\ Sekania for 
useful discussions. 
S.R.M. was supported by a scholarship of the Friedrich-Ebert-Stiftung.
V.M.\ acknowledges support from the Bundesministerium f\"ur Bildung
und Forschung.


\begin{thebibliography}{*}
\bibitem{sizeoffieldamplitude} Close to a critical point, the linear response
regime shrinks and very small amplitudes of the external field are required.
\end{thebibliography}
\end{document}